\documentclass[%
superscriptaddress,
showpacs,preprintnumbers,
amsmath,amssymb,
aps,
prl,
twocolumn
]{revtex4-1}
\usepackage{ulem}
\usepackage{graphicx}
\usepackage{dcolumn}
\usepackage{bm}

\begin {document}

\title{Molecular mechanism of anion permeation through aquaporin 6}

\author{Eiji Yamamoto}
\email{eiji.yamamoto@sd.keio.ac.jp}
\affiliation{%
Department of System Design Engineering, Keio University, Yokohama, Kanagawa 223-8522, Japan
}%

\author{Keehyoung Joo}
\affiliation{%
Center for Advanced Computation, Korea Institute for Advanced Study, Seoul 02455, Korea
}%

\author{Jooyoung Lee}
\affiliation{%
School of Computational Sciences, Korea Institute for Advanced Study, Seoul 02455, Korea
}%

\author{Mark S. P. Sansom}
\affiliation{%
Department of Biochemistry, University of Oxford, South Parks Road, Oxford, OX1 3QU, UK
}%

\author{Masato Yasui}
\affiliation{%
Department of Pharmacology, Keio University School of Medicine, Tokyo 160-8582, Japan
}%



\begin{abstract}
Aquaporins (AQPs) are recognized as transmembrane water channels that facilitate selective water permeation through their monomeric pores. 
Among the AQP family, AQP6 has a unique characteristic as an anion channel, which is allosterically controlled by pH conditions and is eliminated by a single amino acid mutation.
However, the molecular mechanism of anion permeation through AQP6 remains unclear.
Using molecular dynamics simulations in the presence of a transmembrane voltage utilizing an ion concentration gradient, we show that chloride ions permeate through the pore corresponding to the central axis of the AQP6 homotetramer.
Under low pH conditions, a subtle opening of the hydrophobic selective filter (SF), located near the extracellular part of the central pore, becomes wetted and enables anion permeation.
Our simulations also indicate that a single mutation (N63G) in human AQP6, located at the central pore, significantly reduces anion conduction, consistent with experimental data.
Moreover, we demonstrate the pH-sensing mechanism in which the protonation of H184 and H189 under low pH conditions allosterically triggers the gating of the SF region.
These results suggest a unique pH-dependent allosteric anion permeation mechanism in AQP6 and could clarify the role of the central pore in some of the AQP tetramers.
\end{abstract}

\maketitle

Aquaporins (AQPs) are generally known as transmembrane proteins that facilitate selective permeation of water molecules.
In mammalian cells, thirteen isoforms of AQPs (AQP0-12) have been identified that the each AQP plays specific physiological functions with different tissue distributions~\cite{AgreKingYasuiGugginoOttersenFujiyoshiEngelNielsen2002, VerkmanAndersonPapadopoulos2014}.
While most AQPs facilitate the permeation of water molecules, some AQPs, such as aquaglyceroporins, are known to permeate other small molecules~\cite{HubGroot2008, WagnerUngerSalmanKitchenBillYool2022}, e.g., glycerol~\cite{FuLibsonMierckeWeitzmanNollertKrucinskiStroud2000}, urea~\cite{LitmanSogaardZeuthen2009}, hydrogen peroxide~\cite{MillerDickinsonChang2010}, gas~\cite{NakhoulDavisRomeroBoron1998, WangCohenBoronSchultenTajkhorshid2007}, anions~\cite{YasuiHazamaKwonNielsenGugginoAgre1999}, cations~\cite{YoolCampbell2012, HendersonNakayamaWhitelawBruningAndersonTyermanRameshMartinacYool2023}, etc.
The structure of AQPs, which selectively allow the permeation of water molecules, has been well-resolved through experimental studies~\cite{MurataMitsuokaHiraiWalzAgreHeymannEngelFujiyoshi2000, SuiHanLeeWalianJap2001}.
Water molecules exhibit single-file diffusion through the tetrameric pore~\cite{ErikssonFischerFriemannEnkaviTajkhorshidNeutze2013}.
Molecular dynamics (MD) simulations have been conducted to elucidate the dynamics of these water molecules~\cite{JensenTajkhorshidSchulten2003, HubAponte-SantamariaGrubmullerGroot2010, YamamotoAkimotoHiranoYasuiYasuoka2014}.
Additionally, a central pore is formed in the center of the homotetramer.
It has been suggested, based on a single mutagenetic experiment, that this central pore may allow the permeation of ions~\cite{HendersonNakayamaWhitelawBruningAndersonTyermanRameshMartinacYool2023}.

AQP6 is recognized as an anion channel with limited water permeability, even though its sequence closely resembles water channels such as AQP0, AQP2, and AQP5~\cite{YasuiHazamaKwonNielsenGugginoAgre1999}.
AQP6 is localized in the intracellular vesicle membrane of epithelial cells, found in kidney~\cite{MaYangKuoVerkman1996, YasuiKwonKnepperNielsenAgre1999}, vaginal~\cite{KimOhLeeAhnKimPark2011}, and benign ovarian tumors~\cite{MaZhouYangDingZhuChen2016}.
Alongside the H$^+$-ATPase transmembrane proton pump, AQP6 exhibits increased anion conductance under a pH range of 4.0 to 5.5, which is the same as that inside an intracellular vesicle~\cite{YasuiHazamaKwonNielsenGugginoAgre1999}.
It might play a role in acid-base homeostasis though the exact physiological roles of AQP6 remain enigmatic~\cite{Michalek2016, RibeiroAlvesYesteChoCalamitaOliveira2021}.

A single amino acid mutation in AQP6 has been demonstrated to change its function from an anion channel to a water channel.
Specifically, the N60G mutation in rat AQP6, corresponding to N63 in human AQP6, eliminates anion permeation and enhances water permeation~\cite{IkedaBeitzKozonoGugginoAgreYasui2002, LiuKozonoKatoAgreHazamaYasui2005}.
This amino acid residue is conserved as a glycine residue in other human AQPs.
However, the molecular mechanism behind anion permeation through AQP6 and the way in which its protonation state, influenced by pH conditions, regulates anion permeation remain unclear~\cite{YasuiHazamaKwonNielsenGugginoAgre1999}.
Here, we elucidate a molecular mechanism of anion permeation through AQP6 using MD simulations.
We show that chloride ions are permeated through a tetrameric central pore in a pH dependent manner.
The central pore of AQP6 is similar to those in various types of ion channels, which are typically formed by multi-subunit assemblies comprised of tetrameric or pentameric subunits.
We find that wetting of the central pore under low pH conditions is crucial for chloride ion permeation.

\begin{figure*}[tb]
\centering
\includegraphics[width = 150 mm,bb= 0 0 517 310]{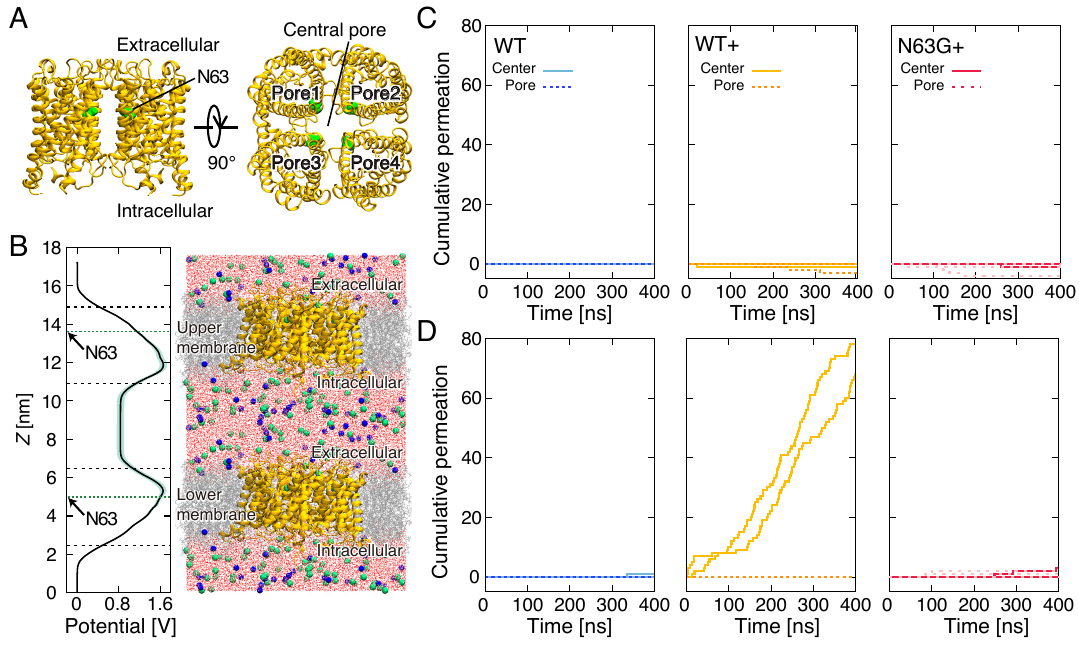}
\caption{Chloride ion permeation through AQP6.
(A)~AQP6 homotetramer.
A central pore is formed at the center of the tetramer.
N63 residues are shown in green.
(B)~CompEL simulation system.
The AQP6 tetramer is embedded in both the upper and lower membranes in the same orientation.
The time-averaged electrostatic potential along the $z$-axis for the WT system is shown with standard deviation (in thin green color).
The electric potential difference across the membrane is generated by the difference in ionic concentrations between the two aqueous compartments.
Horizontal lines represent the boundaries between the membrane and water, and the positions of N63.
(C)~Cumulative chloride ion permeation through the monomeric pores (dashed lines) and central pore (solid lines) in AQP6 tetramers within the upper and (D)~lower membranes ($2 \times 400 \, \mathrm{ns}$ simulations): (left) WT under neutral pH condition, (middle) WT under low pH condition (WT+), and (right) N63G under low pH condition (N63G+).}
\label{fig1}
\end{figure*}

\begin{figure*}[tb]
\centering
\includegraphics[width = 150 mm,bb= 0 0 484 440]{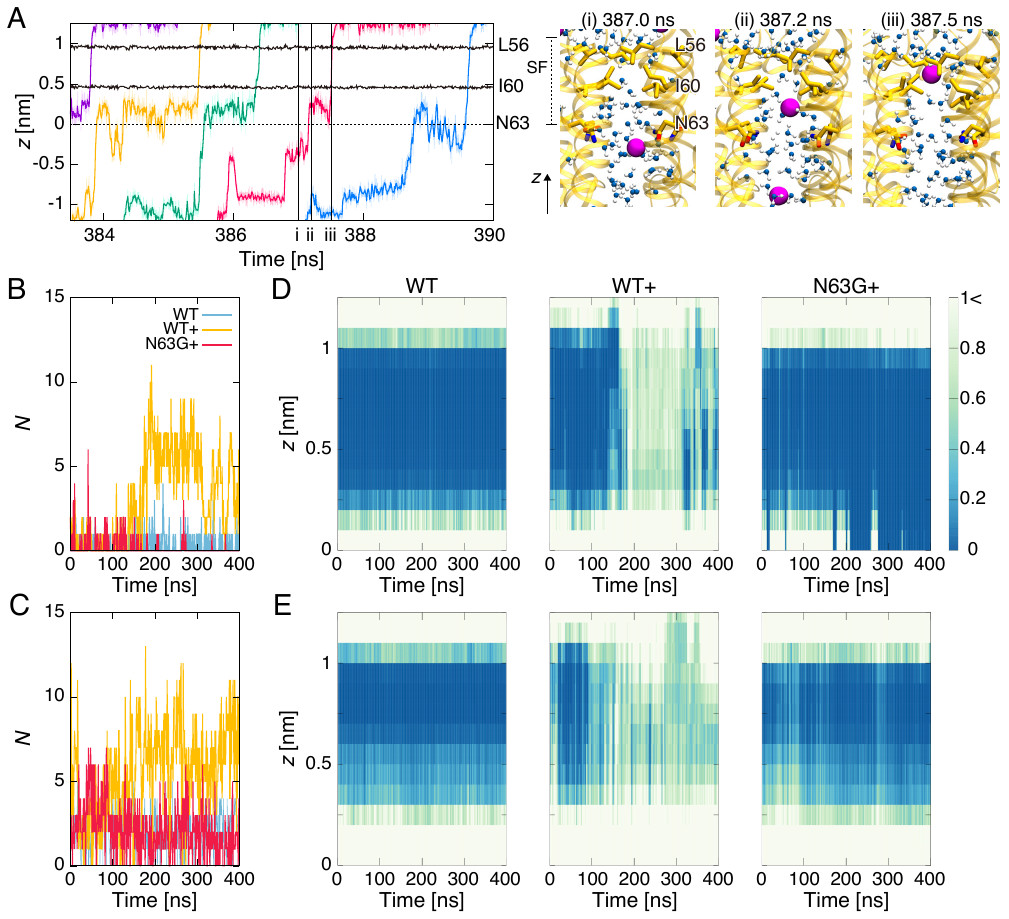}
\caption{Chloride ion permeation and wetting of the SF region in the central pore.
(A)~Trajectories of chloride ions within the central pore of WT AQP6 tetramer under low pH conditions.
$z = 0 \, \mathrm{nm}$ corresponds to  the position of N63 (black dashed line), and the positions of L56 and I60 are shown with black solid lines.
Trajectories averaged over $10 \, \mathrm{ps}$ intervals are shown with solid lines, while the original trajectories saved at $1 \, \mathrm{ps}$ intervals are indicated by thin lines.
Each color represents a different ion.
Representative snapshots of the selective filter with water molecules and chloride ions at each of the time points (i)--(iii) are shown.
AQP6, chloride ion, and water molecules are colored yellow, magenta, and blue, respectively.
L56, I60, and N63 constitute the SF region.
(B)~Number of water molecules within the SF region of AQP6 in the upper and (C)~lower membranes.
Different colored lines represent WT under neutral pH, WT under low pH (WT+), and N63G under low pH conditions (N63G+).
(D)~Wetting state heatmap of the SF region of AQP6 in the upper and (E)~lower membranes.
The color bar represents the number of water molecules within the bin size of $dz=0.1 \, \mathrm{nm}$ along the $z$-axis.
All values above 1 are colored with light green.}
\label{fig2}
\end{figure*}

\begin{figure*}[tb]
\centering
\includegraphics[width = 160 mm,bb= 0 0 639 442]{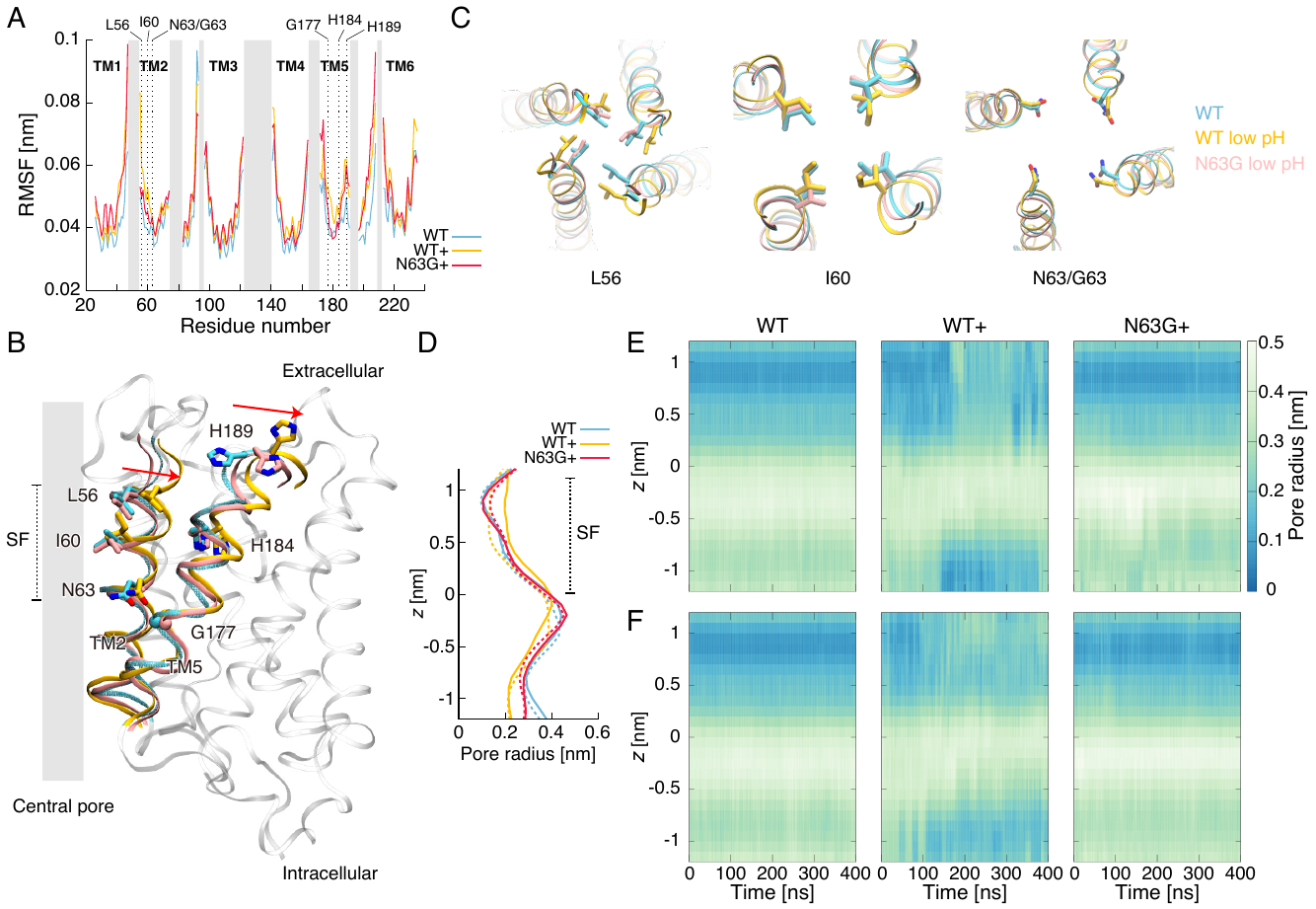}
\caption{Gating of the central pore in AQP6 tetramer for anion permeation in the lower membrane (see Fig.~\ref{fig1}).
(A)~Comparison of RMSF values for transmembrane helices in WT, WT+, and N63G+.
(B)~Comparison of monomer conformations.
Representative structures were obtained from CompEL simulations and were fitted to that of WT.
The transmembrane helices TM2 and TM5 are shown in the same color in panels (A) and (C).
Helices colored white are other helices from the WT.
(C)~SF region of the central pore viewed from the extracellular side.
(D)~Pore radius profile along the central pore.
The dashed and solid lines represent the pore radius in the upper and lower membranes, respectively.
$z=0 \, \mathrm{nm}$ corresponds to the position of N63 (the same as Fig.~\ref{fig2}).
The data are averaged over the last $100 \, \mathrm{ns}$ from two separate runs.
(E)~Heatmap of the central pore radius variation in the AQP6 tetramer over time and $z$ position in the upper and (F)~lower membranes.}
\label{fig3}
\end{figure*}

\begin{figure}[tb]
\centering
\includegraphics[width = 75 mm,bb= 0 0 246 351]{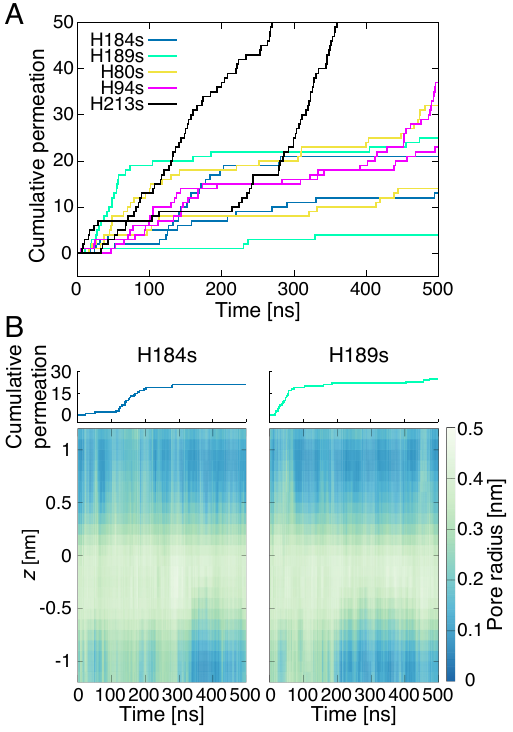}
\caption{Histidine protonation under low pH conditions.
(A)~Cumulative permeation of chloride ions through AQP6 in the lower membrane (see Fig.~\ref{fig1}B).
All histidine residues are double-protonated except for a selected histidine, which is single-protonated.
For example, H184s indicates that H184 is in a single-protonated state.
$2 \times 500 \, \mathrm{ns}$ simulations were performed for each condition.
(B)~Cumulative permeation of chloride ions and the corresponding heatmap of the central pore radius variation in the AQP6 tetramer over time and $z$ position in the lower membrane.
Only H184 (left) or H189 (right) was maintained in a pH-neutral condition.
The structure at time $0 \, \mathrm{ns}$ is the equilibrated structure under low pH conditions.}
\label{fig4}
\end{figure}

\section*{Results}

\subsection*{Homology model of AQP6}
Homology modeling of human AQP6 was performed because its structure has not yet been determined experimentally (see Fig.~\ref{fig1}A and Methods for details). 
The modeling protocol identified AQP2 (PDB: 4NEF)~\cite{FrickErikssonMattiaObergHedfalkNeutzeWillemDeenTornroth-Horsefield2014}, AQP5 (PDB: 3D9S)~\cite{HorsefieldNordenFellertBackmarkTornroth-HorsefieldScheltingaKvassmanKjellbomJohansonNeutze2008}, SoPIP2;1 (PDB: 3CN5)~\cite{NyblomFrickWangEkvallHallgrenHedfalkNeutzeTajkhorshidTornroth-Horsefield2009}, and AtTIP2;1 (PDB: 5I32)~\cite{KirschtKaptanBienertChaumontNissenGrootKjellbomGourdonJohanson2016} as template structures for AQP6. 
To evaluate the stability of the modeled protein conformation within the lipid bilayer, MD simulations of AQP6 tetramer embedded in a lipid bilayer were performed for $1 \, \mathrm{\mu s}$.
The root mean square deviations (RMSDs) for each AQP6 monomer remained within $0.3 \, \mathrm{nm}$ and were stable during the simulations, i.e. the protein conformations were stable within the lipid bilayer (see Fig.~S1).
In the MD simulations, filling of the conventional each monomeric pore with water molecules was observed, suggesting that the homology model of AQP6 is feasible.

\subsection*{Anion permeation through the tetrameric central pore}
To investigate the mechanism of anion permeation, we employed the computational electrophysiology (CompEL) method~\cite{KutznerGrubmullerGrootZachariae2011} in a system where two membranes separate the aqueous compartments in periodic boundary conditions (see Fig.~\ref{fig1}B and details in methods).
This method has been used for several membrane channel systems~\cite{HubAponte-SantamariaGrubmullerGroot2010, KutznerKoepferMachtensGrootSongZachariae2016} to investigate the molecular mechanism of ion permeations.
The electric potential difference across the membrane was induced by an ionic concentration gradient between the two aqueous compartments.
By orienting the proteins in the same direction in each membrane, both cases were simulated: one with high (i.e. positive) electric potential on the extracellular side and another with a high (positive) electric potential on the intracellular side.

We performed two sets of $400 \, \mathrm{ns}$ CompEL simulations for each of three different systems: wild type (WT) in neutral pH conditions, WT in low pH conditions (WT+), and N63G in low pH conditions (N63G+).
Here, in order to observe sufficient anion permeation events, we put 16e$^-$ charge imbalance, generating $0.8 \, \mathrm{V}$~\cite{HubAponte-SantamariaGrubmullerGroot2010, MachtensKortzakLanscheLeinenweberKilianBegemannZachariaeEwersGrootBrionesFahlke2015, KutznerKoepferMachtensGrootSongZachariae2016, KlesseTuckerSansom2020}.
Experimental studies have shown that anions permeate through AQP6 under a pH lower than 5.5~\cite{YasuiHazamaKwonNielsenGugginoAgre1999}.
In our simulations, more than 60 chloride ions were permeated through the central pore of the AQP6 tetramer in WT+, whereas only a few chloride ions were permeated during the two $400 \, \mathrm{ns}$ simulation in WT (see Fig.~\ref{fig1}CD).
Interestingly, ion permeation was observed solely in the lower membrane during the CompEL simulation.
This is the same situation that anions are abundant in the intracellular side and are conducted through the vesicle from intracellular to extracellular compartment.
These results are in good agreement with experimentally observed slight outward rectification~\cite{YasuiHazamaKwonNielsenGugginoAgre1999, LiuKozonoKatoAgreHazamaYasui2005}.

Moreover, our simulations showed that the N63G mutation in AQP6 eliminates chloride ion permeation under low pH conditions (N63G+), which is consistent with experimental observations~\cite{LiuKozonoKatoAgreHazamaYasui2005} (see Fig.~\ref{fig1}CD).
Even with an $0.8 \, \mathrm{V}$ voltage difference across the membrane, which is five times higher than physiological values (and is close to the breakdown voltage of the lipid bilayer), the rate of chloride ion permeation under low pH conditions was significantly reduced due to the single N63G mutation.
N63 is located in the middle of the central pore of the AQP6 tetramer and may work as a component of the SF.

\subsection*{Wetting of the selective filter located in the central pore is crucial for the anion permeation}
Figure~\ref{fig2}A shows the time series of chloride ions passing through the central pore under low pH conditions (see Movie~S1). 
The amino acids L56, I60, and N63 are key components that constitute the SF in the central pore.
Chloride ions entering from the intracellular side initially (Fig.~\ref{fig2}A, snapshot (i) at $387.0 \, \mathrm{ns}$) interact with water molecules and an `entry site' in the intracellular ($z < 0$) half of the pore formed by an N63 side chain.
The chloride ion then jumps to a `filter binding site' (snapshot (ii) at $387.2 \, \mathrm{ns}$) where it interacts with water molecules and the hydrophobic side chains of I60, before moving to an `exit site' (snapshot (iii) at $387.5 \, \mathrm{ns}$) formed by the hydrophobic L56 side chains, and then leaves via the extracellular mouth of the pore.
As seen in the snapshots in Fig.~\ref{fig2}A, the region below I60 in the central pore is wetted by several water molecules.
In contrast, the narrow region between L56 and I60, where chloride ions pass through quickly, contains a few water molecules.
For the permeation of chloride ions through this narrow region, the narrow region itself must be wetted by some water molecules whilst the chloride ion also interacts with hydrophobic side chains.
This phenomenon is also observed in ion permeation through other ion channels~\cite{LynchRaoSansom2020}.
We note that chloride ion interaction with hydrophobic side chains in addition to waters has also been suggested in a number of other anion channels~\cite{KlesseRaoTuckerSansom2020, JojoaCruzSaotomeTsuiLeeSansomMurthyPatapoutianWard2022} and transporters~\cite{PhanChamorroMartinezSearaCrainSansomTucker2023}.
In the case of AQP6, chloride ions are permeated through the wetted SF region in a hydrated state.

The number of water molecules in the SF region between $0.2 \, \mathrm{nm} < z < 1.0 \, \mathrm{nm}$ was calculated, where $z = 0 \, \mathrm{nm}$ corresponds to the position of N63.
Only a few water molecules are present in this region for both WT and N63G+, while for WT+, 5 to 10 water molecules are in the region (see Fig.~\ref{fig2}BC).

Figures~\ref{fig2}DE show the heatmap of the temporal variation in wetting states in the SF region.
The permeation of chloride ions is synchronously linked to the time variation of wetting states.
For WT+ in the lower membrane, the SF region remains in a dewetted state between 50 and $120 \, \mathrm{ns}$, transitioning to a wetted state after $120 \, \mathrm{ns}$ (Fig.~\ref{fig2}E).
When WT+ is in its dewetted state, chloride ions are not permeated.
However, AQP6 begins to facilitate ion permeation upon transitioning to a wetted state within the SF region (Fig.~\ref{fig1}D).
In the case of N63G+, the SF region is dehydrated.
N63 has a polar, uncharged side chain where the nitrogen atom forms a hydrogen bond with water molecules.
This is the molecular mechanism by which the N63G mutation eliminates chloride ion permeation.
Thus, the loss of the N63 side chain removes the `entry site' for chloride ion, which in turn prevents the anion from reaching the `filter site'.
This means that the I60/L56 region of the pore remains narrower and dehydrated.

Differences in the direction of the applied electric field cause variations in the wetting state within the SF region. 
For WT and N63G+, when the extracellular side has a higher electric potential than the intracellular side as seen in the lower membrane in the system (Fig.~\ref{fig1}B), the SF region contains only a few numbers of water molecules, up to around the position of I60 ($z =0.5 \, \mathrm{nm}$) (Figs.~\ref{fig2}CE).
This is not the case when the direction of the electric field is opposite (Figs.~\ref{fig2}BD).
In contrast, for WT+, although the SF region in both the upper and lower membranes is fully hydrated, the degree of hydration in the lower leaflet (Fig~\ref{fig2}E) is higher than that in the upper leaflet (Fig~\ref{fig2}D).
This difference is attributed to variations in the electric field applied to the protein along the $z$-axis (Fig.~\ref{fig1}B) and in particular in the absence (upper leaflet) vs. presence (lower leaflet) of a positive electrostatic potential in the vicinity of the (protonated) H184 side chain is thought to be related to the outward rectification.

\subsection*{Structural mechanism of anion permeability}
To clarify how AQP6 regulates anion permeation, we analyzed its conformational properties.
Figure~\ref{fig3}A shows the root mean square fluctuations (RMSFs) of the C$\alpha$ atoms for each residue in AQP6, focusing on the ion-permeable state when the extracellular electric potential is higher than the intracellular potential (lower membrane in Fig.~\ref{fig1}B).
The RMSFs were calculated for each monomer after fitting it to the homology model, which was used as the reference structure, and then averaged over simulations with a total time of $200 \, \mathrm{ns}$ (utilizing the last $100 \, \mathrm{ns}$ from two separate runs).
The high value of RMSF indicates that the structure exhibits large variations from the reference structure.
The RMSF values in transmembrane helices TM2 and TM5 show a relatively small difference, $\sim 0.01$--$0.02 \, \mathrm{nm}$, between the chloride-permeable structure of WT+ and the non-permeable structures of WT and N63G+ (Fig.~\ref{fig3}A).
The outward movement of TM5, induced by the protonation of H184 and H189, triggers the outward movement of TM2 constituting the SF region (Fig.~\ref{fig3}B).
Since there are no amino acid residues adjacent to H184 and H189 capable of forming a pH-dependent salt bridge, the outward movement of TM5 under low pH is thought to be due to the transmembrane voltage generated by the ionic concentration gradient.

The movement of each monomer contributes to an $\sim 0.1 \, \mathrm{nm}$ opening of the SF region (Figs.~\ref{fig3}CD).
The pore size of the SF region in WT+, constituted by L56, I60, and N63 within TM2, is larger than that in WT and N63G.
Figures~\ref{fig3}EF show the heatmap of the temporal variation in the central pore radius.
Because all simulations started with the WT structure under neutral pH conditions, the radius of the SF region initially had a small pore radius of $0.1 \, \mathrm{nm}$.
In the case of WT+ (lower membrane) after $100 \, \mathrm{ns}$, the SF transitions to an open state with a radius of $0.2 \, \mathrm{nm}$, while the radii of WT and N63G+ remain closed (Figs.~\ref{fig3}DF).

\subsection*{Protonation of histidine residues}
AQP6 is abundant in intracellular vesicles, colocalizing with H$^+$-ATPase.
Thus, a low pH inside the intracellular vesicles may be the natural activator of AQP6, and it is reasonable to assume that AQP6 has pH sensors near its extracellular side. 
To determine which histidine residues are essential for anion permeation, we performed $2 \times 500 \, \mathrm{ns}$ CompEL simulations.
In these simulations, a selected histidine residue was maintained in a pH-neutral state (single protonated state), while the other histidine residues were double protonated.
If the selected histidine residue is crucial for pH sensing, AQP6 does not permeate chloride ions; otherwise, it does.
Figure~\ref{fig4}A shows the cumulative permeation of chloride ions in the lower membrane. 
Because the initial AQP6 structure was obtained from simulations of the WT+, chloride ions were permeated at the beginning of the simulation.
After several nanoseconds, chloride ion permeation through AQP6 significantly decreases, except for H213 in its single protonated state (H213s).
Figure~\ref{fig4}B shows the heatmap of the temporal variation in the central pore radius of AQP6 tetramer in H184s and H189s simulations (also see Fig.~S2 for the pore radius profile along the central pore).
The number of chloride ion permeations and the pore radius are synchronized, where cumulative permeation shows a plateau corresponding to a decrease in the pore radius.

H184 and H189 are located on the extracellular side, while H80 and H94 are on the intracellular side (Fig.~S3).
H184, an amino acid residue in the ar/R region~\cite{FujiyoshiMitsuokaGrootPhilippsenGrubmullerAgreEngel2002}, is conserved across various species of AQP6, suggesting an important role for this specific residue.
Residue 189 can be either histidine or tyrosine, depending on the species, and these residues have similar side chains, i.e. the pH sensitivity mediated by residue 189 might vary depending on the species.
In the CompEL simulations with two different initial conditions, AQP6 with either single protonated states of H80 (H80s) or H94 (H94s) showed chloride ion permeation after $400 \, \mathrm{ns}$ (see Fig.~\ref{fig4}A).
These results suggest that double protonation of H80 and H94 might not be critical for chloride ion permeation.
Thus, H184 and H189 at the extracellular end of TM5 are thought to regulate anion permeations depending on pH conditions.

\section*{Discussion}
In summary, we have performed MD simulations to investigate the molecular mechanisms of anion permeation through AQP6.
We found that chloride ions permeate through the central pore of the AQP6 tetramer, rather teliminatedhan through each monomeric pore.
Anion permeation was activated under low pH conditions and was eliminated by the N63G mutation.
These phenomena are consistent with previous experimental results~\cite{YasuiHazamaKwonNielsenGugginoAgre1999, LiuKozonoKatoAgreHazamaYasui2005}.
It is worth noting that the hAQP5 structure (3D9S), in which a lipid tail is blocking the intracellular half of the central pore~\cite{HorsefieldNordenFellertBackmarkTornroth-HorsefieldScheltingaKvassmanKjellbomJohansonNeutze2008}.
This reinforces our suggestion that the hydrophobic central pore is crucial for anion permeation.
Since the structure of AQP6 has not been solved experimentally, we used a homology-modeled structure of AQP6.
MD simulations showed that the predicted AQP6 structure did not change significantly within the lipid membrane.
Given the qualitative agreement between our simulations and previous experimental results, the MD simulation protocol using computationally predicted protein structures could serve as a powerful tool for investigating the dynamics of proteins whose experimental structures are currently not available.

Moreover, we have revealed the pH-sensing mechanism and water-mediated anion permeation through the SF region of the central pore.
The protonation of the H184 and H189 residues in the extracellular part induces an outward movement of TM5, which allosterically triggers the subtle opening and wetting of the SF regions.
 A variety of ion channels are expressed in biological membranes and play key roles in signaling.
The structure of the central pore in tetrameric AQP6 resembles the multi-subunit assemblies found in other ion channels, where usually tetrameric or pentameric subunits form a pore~\cite{MaffeoBhattacharyaYooWellsAksimentiev2012}.
It is also recognized that acid-sensing ion channels and proton-gated ion channels exhibit pH-dependent regulation of ion permeation.
Thus, our findings are important for providing a microscopic physical rationale for the fundamental gating and wetting mechanisms of the SF region~\cite{LynchRaoSansom2020}, which are common in other ion channels.

In this study, we applied a  voltage difference of $0.8 \, \mathrm{V}$ (16e$^-$ charge imbalance) across the membrane, which is higher than the physiological values of $\sim 0.05$--$0.2 \, \mathrm{V}$.
However, when we applied only a half-charge imbalance, ion permeation was insufficiently observed within the timescale of this study.
It is also known that the wetting of the hydrophobic SF region in other channels occurs at higher voltages in previous MD simulation studies~\cite{HubAponte-SantamariaGrubmullerGroot2010, MachtensKortzakLanscheLeinenweberKilianBegemannZachariaeEwersGrootBrionesFahlke2015, KutznerKoepferMachtensGrootSongZachariae2016, TrickChelvaniththilanKlesseAryalWallaceTuckerSansom2016, KlesseTuckerSansom2020, LynchKlesseRaoTuckerSansom2021}.
For further investigation into the detailed mechanisms of wetting and dewetting processes, the use of alternative force fields is one option.
The polarizable water model increases the wettability of the hydrophobic region of the pore~\cite{LynchKlesseRaoTuckerSansom2021}.
For more physiologically relevant situations, asymmetric membranes could affect the partial transmembrane voltage.
Examination of other types of anion permeation~\cite{IkedaBeitzKozonoGugginoAgreYasui2002} could provide a more comprehensive understanding of anion permeation mechanisms.
We speculate that these results will aid in the design of biomimetic nanopores~\cite{RaoKlesseStansfeldTuckerSansom2019, AraiYamamotoKoishiHiranoYasuokaEbisuzaki2023}.

\section{methods}
\subsection*{Homology modeling of AQP6 structure}
The protein model structure of Homo sapiens AQP6 (residues 14-249) was generated using a template-based modeling protocol~\cite{JooJoungLeeKimChengManavalanJoungHeoLeeNamLeeLeeLee2016}.
The protocol is based on global optimization~\cite{JooLeeKimLeeLee2008, JooLeeSeoLeeKimLee2009} to generate 3D protein models and has been quite successful in CASP protein structure prediction experiments.
This protocol relies on global optimization~\cite{JooLeeKimLeeLee2008, JooLeeSeoLeeKimLee2009} to generate 3D protein models and identified template structures, 3CN5, 4NEF, 3D9S, and 5I32 for AQP6. 
Using these templates and target sequences, a multiple sequence alignment (MSA) was carried out using the MSACSA method~\cite{JooLeeKimLeeLee2008}.
The resultant sequence similarity ranged between $33 \, \mathrm{\%}$ and $65 \, \mathrm{\%}$.
With the resultant MSA and templates, a 3D protein model was generated using the global optimization method of conformational space of annealing~\cite{JooLeeKimLeeLee2008, JooLeeSeoLeeKimLee2009, LeeScheragaRackovsky1997}.
The side chains of the models were refined by rotamer optimization and MD simulation~\cite{JooJoungLeeKimChengManavalanJoungHeoLeeNamLeeLeeLee2016}. 

\subsection*{MD simulations of AQP6: single membrane system}
For MD simulations, a homotetrameric assembly of AQP6 was embedded in a lipid bilayer comprising 300 1-palmitoyl-2-oleoyl-glycero-3-phosphocholine (POPC) molecules.
The simulation system was solvated with approximately 21,000 TIP3P water molecules, and NaCl ions were added at a concentration of 150 mM to neutralize the system.
We performed MD simulations on four different systems to study the behavior of AQP6; the wild-type AQP6 (WT) under neutral pH conditions, the wild-type AQP6 (WT+) under low pH conditions, the N63G-mutated AQP6 (N63G) under neutral pH conditions, and the N63G-mutated AQP6 (N63G+) under low pH conditions.
Under low pH conditions, assuming the same pH range of 4.0 to 5.5 as inside an intracellular vesicle~\cite{YasuiHazamaKwonNielsenGugginoAgre1999}, all histidine residues were protonated at both nitrogen sites in their imidazole rings.
Each system was simulated for a time duration of $1 \, \mathrm{\mu s}$ with a time step of $2 \, \mathrm{fs}$.
The systems were subjected to pressure scaling to $0.1 \, \mathrm{MPa}$ using a Parrinello-Rahman barostat~\cite{ParrinelloRahman1981}, and temperature scaling to $310 \, \mathrm{K}$ using a v-rescale thermostat~\cite{BussiZykova-TimanParrinello2009}.
Hydrogen bond lengths were constrained to their equilibrium values using the LINCS method~\cite{HessBekkerBerendsenFraaije1997}.
Force fields used were AMBERff99SB-ildn~\cite{Lindorff-LarsenPianaPalmoMaragakisKlepeisDrorShaw2010} for protein, Slipid~\cite{JambeckLyubartsev2012a} for POPC, and TIP3P water models~\cite{JorgensenChandrasekharMaduraImpeyKlein1983}.
The particle mesh Ewald method~\cite{EssmannPereraBerkowitzDardenLeePedersen1995} was used with a specified direct space cutoff distance of $1.2 \, \mathrm{nm}$. 
All simulations were performed using GROMACS 2018.

\subsection*{MD simulations of AQP6 under ion concentration gradients: double membrane system}
To induce ion flux through AQP6, we used the CompEL protocol~\cite{KutznerGrubmullerGrootZachariae2011}.
In periodic boundary conditions, two membranes were placed to separate two compartments of solution, generating an ion concentration difference across the membrane (see Fig.~\ref{fig1}).
In the two-compartment system neutralized with $150 \, \mathrm{mM}$ NaCl ions, one compartment had 8 extra sodium ions and the other had 8 extra chloride ions to create an electrostatic potential difference.
Pre-equilibrated simulations ($ 1 \, \mathrm{\mu s}$) of the single membrane system were duplicated to generate the double membrane system for the CompEL simulations.
Two sets of $ 0.4 \, \mathrm{\mu s}$ simulations were carried out for three different systems: WT, WT+, and N63G+, using the same representations as in the single membrane systems.

To identify the pH sensing histidine residue, we performed simulations in which a selected histidine residue was maintained in a pH-neutral state (single protonated state), while all other histidine residues were double protonated.
These simulations were carried out for two sets of $ 0.5 \, \mathrm{\mu s}$ for single protonated states of H80, H94, H184, H189, and H213, respectively (see Fig.~S3).
For the initial configuration, a pre-equilibrated simulation system ($ 1 \, \mathrm{\mu s}$) of WT+ (single membrane system) was used for all simulations.

In CompEL simulations, if an ion permeated into a different compartment via AQP6, the positions of the ion and water were exchanged using Monte Carlo criteria to maintain the overall ionic concentration difference.
The electrostatic potential was calculated using the Poisson equation with the implemented gmx potential commands in GROMACS.
A non-physical offset in the electrostatic potential between the top and bottom of the simulation box was corrected using a linear adjustment, taking into account the periodic boundary conditions~\cite{HubAponte-SantamariaGrubmullerGroot2010}.

\subsection*{Acknowledgments}
We thank Dr.~Takahisa Maki, Dr.~Masayuki Iwamoto, and Dr.~Shigetoshi Oiki for fruitful discussion.
This work was supported by KAKENHI Grant-in-Aid (No. 18K13517) from JSPS, Keio University Research Grant for Young Researcher's Program, and Basic Science Research Program through the National Research Foundation of Korea (NRF) funded by the Ministry of Science and ICT [NRF-2017R1E1A1A01077717]. We thank Korea Institute for Advanced Study for providing computing resources for this work.


\begin{thebibliography}{55}%
\makeatletter
\providecommand \@ifxundefined [1]{%
 \@ifx{#1\undefined}
}%
\providecommand \@ifnum [1]{%
 \ifnum #1\expandafter \@firstoftwo
 \else \expandafter \@secondoftwo
 \fi
}%
\providecommand \@ifx [1]{%
 \ifx #1\expandafter \@firstoftwo
 \else \expandafter \@secondoftwo
 \fi
}%
\providecommand \natexlab [1]{#1}%
%
\providecommand \bibnamefont  [1]{#1}%
\providecommand \bibfnamefont [1]{#1}%
\providecommand \citenamefont [1]{#1}%
\providecommand \href@noop [0]{\@secondoftwo}%
\providecommand \href [0]{\begingroup \@sanitize@url \@href}%
\providecommand \@href[1]{\@@startlink{#1}\@@href}%
\providecommand \@@href[1]{\endgroup#1\@@endlink}%
\providecommand \@sanitize@url [0]{\catcode `\\12\catcode `\$12\catcode
  `\&12\catcode `\#12\catcode `\^12\catcode `\_12\catcode `\%12\relax}%
\providecommand \@@startlink[1]{}%
\providecommand \@@endlink[0]{}%
\providecommand \url  [0]{\begingroup\@sanitize@url \@url }%
\providecommand \@url [1]{\endgroup\@href {#1}{\urlprefix }}%
\providecommand \urlprefix  [0]{URL }%
%
\providecommand \doibase [0]{http://dx.doi.org/}%
\providecommand \selectlanguage [0]{\@gobble}%
\providecommand \bibinfo  [0]{\@secondoftwo}%
\providecommand \bibfield  [0]{\@secondoftwo}%
%
\providecommand \BibitemOpen [0]{}%
%
%
%
\providecommand \BibitemShut  [1]{\csname bibitem#1\endcsname}%
\let\auto@bib@innerbib\@empty
\bibitem [{\citenamefont {Agre}\ \emph {et~al.}(2002)\citenamefont {Agre},
  \citenamefont {King}, \citenamefont {Yasui}, \citenamefont {Guggino},
  \citenamefont {Ottersen}, \citenamefont {Fujiyoshi}, \citenamefont {Engel},\
  and\ \citenamefont
  {Nielsen}}]{AgreKingYasuiGugginoOttersenFujiyoshiEngelNielsen2002}%
  \BibitemOpen
  \bibfield  {author} {\bibinfo {author} {\bibfnamefont {P.}~\bibnamefont
  {Agre}}, \bibinfo {author} {\bibfnamefont {L.~S.}\ \bibnamefont {King}},
  \bibinfo {author} {\bibfnamefont {M.}~\bibnamefont {Yasui}}, \bibinfo
  {author} {\bibfnamefont {W.~B.}\ \bibnamefont {Guggino}}, \bibinfo {author}
  {\bibfnamefont {O.~P.}\ \bibnamefont {Ottersen}}, \bibinfo {author}
  {\bibfnamefont {Y.}~\bibnamefont {Fujiyoshi}}, \bibinfo {author}
  {\bibfnamefont {A.}~\bibnamefont {Engel}}, \ and\ \bibinfo {author}
  {\bibfnamefont {S.}~\bibnamefont {Nielsen}},\ }\href@noop {} {\bibfield
  {journal} {\bibinfo  {journal} {J. Physiol.}\ }\textbf {\bibinfo {volume}
  {542}},\ \bibinfo {pages} {3} (\bibinfo {year} {2002})}\BibitemShut {NoStop}%
\bibitem [{\citenamefont {Verkman}\ \emph {et~al.}(2014)\citenamefont
  {Verkman}, \citenamefont {Anderson},\ and\ \citenamefont
  {Papadopoulos}}]{VerkmanAndersonPapadopoulos2014}%
  \BibitemOpen
  \bibfield  {author} {\bibinfo {author} {\bibfnamefont {A.~S.}\ \bibnamefont
  {Verkman}}, \bibinfo {author} {\bibfnamefont {M.~O.}\ \bibnamefont
  {Anderson}}, \ and\ \bibinfo {author} {\bibfnamefont {M.~C.}\ \bibnamefont
  {Papadopoulos}},\ }\href@noop {} {\bibfield  {journal} {\bibinfo  {journal}
  {Nat. Rev. Drug Discovery}\ }\textbf {\bibinfo {volume} {13}},\ \bibinfo
  {pages} {259} (\bibinfo {year} {2014})}\BibitemShut {NoStop}%
\bibitem [{\citenamefont {Hub}\ and\ \citenamefont
  {de~Groot}(2008)}]{HubGroot2008}%
  \BibitemOpen
  \bibfield  {author} {\bibinfo {author} {\bibfnamefont {J.~S.}\ \bibnamefont
  {Hub}}\ and\ \bibinfo {author} {\bibfnamefont {B.~L.}\ \bibnamefont
  {de~Groot}},\ }\href@noop {} {\bibfield  {journal} {\bibinfo  {journal}
  {Proc. Natl. Acad. Sci. USA}\ }\textbf {\bibinfo {volume} {105}},\ \bibinfo
  {pages} {1198} (\bibinfo {year} {2008})}\BibitemShut {NoStop}%
\bibitem [{\citenamefont {Wagner}\ \emph {et~al.}(2022)\citenamefont {Wagner},
  \citenamefont {Unger}, \citenamefont {Salman}, \citenamefont {Kitchen},
  \citenamefont {Bill},\ and\ \citenamefont
  {Yool}}]{WagnerUngerSalmanKitchenBillYool2022}%
  \BibitemOpen
  \bibfield  {author} {\bibinfo {author} {\bibfnamefont {K.}~\bibnamefont
  {Wagner}}, \bibinfo {author} {\bibfnamefont {L.}~\bibnamefont {Unger}},
  \bibinfo {author} {\bibfnamefont {M.~M.}\ \bibnamefont {Salman}}, \bibinfo
  {author} {\bibfnamefont {P.}~\bibnamefont {Kitchen}}, \bibinfo {author}
  {\bibfnamefont {R.~M.}\ \bibnamefont {Bill}}, \ and\ \bibinfo {author}
  {\bibfnamefont {A.~J.}\ \bibnamefont {Yool}},\ }\href@noop {} {\bibfield
  {journal} {\bibinfo  {journal} {Inter. J. Mol. Sci.}\ }\textbf {\bibinfo
  {volume} {23}},\ \bibinfo {pages} {1388} (\bibinfo {year}
  {2022})}\BibitemShut {NoStop}%
\bibitem [{\citenamefont {Fu}\ \emph {et~al.}(2000)\citenamefont {Fu},
  \citenamefont {Libson}, \citenamefont {Miercke}, \citenamefont {Weitzman},
  \citenamefont {Nollert}, \citenamefont {Krucinski},\ and\ \citenamefont
  {Stroud}}]{FuLibsonMierckeWeitzmanNollertKrucinskiStroud2000}%
  \BibitemOpen
  \bibfield  {author} {\bibinfo {author} {\bibfnamefont {D.}~\bibnamefont
  {Fu}}, \bibinfo {author} {\bibfnamefont {A.}~\bibnamefont {Libson}}, \bibinfo
  {author} {\bibfnamefont {L.~J.~W.}\ \bibnamefont {Miercke}}, \bibinfo
  {author} {\bibfnamefont {C.}~\bibnamefont {Weitzman}}, \bibinfo {author}
  {\bibfnamefont {P.}~\bibnamefont {Nollert}}, \bibinfo {author} {\bibfnamefont
  {J.}~\bibnamefont {Krucinski}}, \ and\ \bibinfo {author} {\bibfnamefont
  {R.~M.}\ \bibnamefont {Stroud}},\ }\href@noop {} {\bibfield  {journal}
  {\bibinfo  {journal} {Science}\ }\textbf {\bibinfo {volume} {290}},\ \bibinfo
  {pages} {481} (\bibinfo {year} {2000})}\BibitemShut {NoStop}%
\bibitem [{\citenamefont {Litman}\ \emph {et~al.}(2009)\citenamefont {Litman},
  \citenamefont {S{\o}gaard},\ and\ \citenamefont
  {Zeuthen}}]{LitmanSogaardZeuthen2009}%
  \BibitemOpen
  \bibfield  {author} {\bibinfo {author} {\bibfnamefont {T.}~\bibnamefont
  {Litman}}, \bibinfo {author} {\bibfnamefont {R.}~\bibnamefont {S{\o}gaard}},
  \ and\ \bibinfo {author} {\bibfnamefont {T.}~\bibnamefont {Zeuthen}},\
  }\href@noop {} {\bibfield  {journal} {\bibinfo  {journal} {Aquaporins}\
  }\textbf {\bibinfo {volume} {190}},\ \bibinfo {pages} {327} (\bibinfo {year}
  {2009})}\BibitemShut {NoStop}%
\bibitem [{\citenamefont {Miller}\ \emph {et~al.}(2010)\citenamefont {Miller},
  \citenamefont {Dickinson},\ and\ \citenamefont
  {Chang}}]{MillerDickinsonChang2010}%
  \BibitemOpen
  \bibfield  {author} {\bibinfo {author} {\bibfnamefont {E.~W.}\ \bibnamefont
  {Miller}}, \bibinfo {author} {\bibfnamefont {B.~C.}\ \bibnamefont
  {Dickinson}}, \ and\ \bibinfo {author} {\bibfnamefont {C.~J.}\ \bibnamefont
  {Chang}},\ }\href@noop {} {\bibfield  {journal} {\bibinfo  {journal} {Proc.
  Natl. Acad. Sci. USA}\ }\textbf {\bibinfo {volume} {107}},\ \bibinfo {pages}
  {15681} (\bibinfo {year} {2010})}\BibitemShut {NoStop}%
\bibitem [{\citenamefont {Nakhoul}\ \emph {et~al.}(1998)\citenamefont
  {Nakhoul}, \citenamefont {Davis}, \citenamefont {Romero},\ and\ \citenamefont
  {Boron}}]{NakhoulDavisRomeroBoron1998}%
  \BibitemOpen
  \bibfield  {author} {\bibinfo {author} {\bibfnamefont {N.~L.}\ \bibnamefont
  {Nakhoul}}, \bibinfo {author} {\bibfnamefont {B.~A.}\ \bibnamefont {Davis}},
  \bibinfo {author} {\bibfnamefont {M.~F.}\ \bibnamefont {Romero}}, \ and\
  \bibinfo {author} {\bibfnamefont {W.~F.}\ \bibnamefont {Boron}},\ }\href@noop
  {} {\bibfield  {journal} {\bibinfo  {journal} {Am. J. Physiol. Cell
  Physiol.}\ }\textbf {\bibinfo {volume} {274}},\ \bibinfo {pages} {C543}
  (\bibinfo {year} {1998})}\BibitemShut {NoStop}%
\bibitem [{\citenamefont {Wang}\ \emph {et~al.}(2007)\citenamefont {Wang},
  \citenamefont {Cohen}, \citenamefont {Boron}, \citenamefont {Schulten},\ and\
  \citenamefont {Tajkhorshid}}]{WangCohenBoronSchultenTajkhorshid2007}%
  \BibitemOpen
  \bibfield  {author} {\bibinfo {author} {\bibfnamefont {Y.}~\bibnamefont
  {Wang}}, \bibinfo {author} {\bibfnamefont {J.}~\bibnamefont {Cohen}},
  \bibinfo {author} {\bibfnamefont {W.~F.}\ \bibnamefont {Boron}}, \bibinfo
  {author} {\bibfnamefont {K.}~\bibnamefont {Schulten}}, \ and\ \bibinfo
  {author} {\bibfnamefont {E.}~\bibnamefont {Tajkhorshid}},\ }\href@noop {}
  {\bibfield  {journal} {\bibinfo  {journal} {J. Struct. Biol.}\ }\textbf
  {\bibinfo {volume} {157}},\ \bibinfo {pages} {534} (\bibinfo {year}
  {2007})}\BibitemShut {NoStop}%
\bibitem [{\citenamefont {Yasui}\ \emph
  {et~al.}(1999{\natexlab{a}})\citenamefont {Yasui}, \citenamefont {Hazama},
  \citenamefont {Kwon}, \citenamefont {Nielsen}, \citenamefont {Guggino},\ and\
  \citenamefont {Agre}}]{YasuiHazamaKwonNielsenGugginoAgre1999}%
  \BibitemOpen
  \bibfield  {author} {\bibinfo {author} {\bibfnamefont {M.}~\bibnamefont
  {Yasui}}, \bibinfo {author} {\bibfnamefont {A.}~\bibnamefont {Hazama}},
  \bibinfo {author} {\bibfnamefont {T.-H.}\ \bibnamefont {Kwon}}, \bibinfo
  {author} {\bibfnamefont {S.}~\bibnamefont {Nielsen}}, \bibinfo {author}
  {\bibfnamefont {W.~B.}\ \bibnamefont {Guggino}}, \ and\ \bibinfo {author}
  {\bibfnamefont {P.}~\bibnamefont {Agre}},\ }\href@noop {} {\bibfield
  {journal} {\bibinfo  {journal} {Nature}\ }\textbf {\bibinfo {volume} {402}},\
  \bibinfo {pages} {184} (\bibinfo {year} {1999}{\natexlab{a}})}\BibitemShut
  {NoStop}%
\bibitem [{\citenamefont {Yool}\ and\ \citenamefont
  {Campbell}(2012)}]{YoolCampbell2012}%
  \BibitemOpen
  \bibfield  {author} {\bibinfo {author} {\bibfnamefont {A.~J.}\ \bibnamefont
  {Yool}}\ and\ \bibinfo {author} {\bibfnamefont {E.~M.}\ \bibnamefont
  {Campbell}},\ }\href@noop {} {\bibfield  {journal} {\bibinfo  {journal} {Mol.
  Aspect. Med.}\ }\textbf {\bibinfo {volume} {33}},\ \bibinfo {pages} {553}
  (\bibinfo {year} {2012})}\BibitemShut {NoStop}%
\bibitem [{\citenamefont {Henderson}\ \emph {et~al.}(2023)\citenamefont
  {Henderson}, \citenamefont {Nakayama}, \citenamefont {Whitelaw},
  \citenamefont {Bruning}, \citenamefont {Anderson}, \citenamefont {Tyerman},
  \citenamefont {Ramesh}, \citenamefont {Martinac},\ and\ \citenamefont
  {Yool}}]{HendersonNakayamaWhitelawBruningAndersonTyermanRameshMartinacYool2023}%
  \BibitemOpen
  \bibfield  {author} {\bibinfo {author} {\bibfnamefont {S.~W.}\ \bibnamefont
  {Henderson}}, \bibinfo {author} {\bibfnamefont {Y.}~\bibnamefont {Nakayama}},
  \bibinfo {author} {\bibfnamefont {M.~L.}\ \bibnamefont {Whitelaw}}, \bibinfo
  {author} {\bibfnamefont {J.~B.}\ \bibnamefont {Bruning}}, \bibinfo {author}
  {\bibfnamefont {P.~A.}\ \bibnamefont {Anderson}}, \bibinfo {author}
  {\bibfnamefont {S.~D.}\ \bibnamefont {Tyerman}}, \bibinfo {author}
  {\bibfnamefont {S.~A.}\ \bibnamefont {Ramesh}}, \bibinfo {author}
  {\bibfnamefont {B.}~\bibnamefont {Martinac}}, \ and\ \bibinfo {author}
  {\bibfnamefont {A.~J.}\ \bibnamefont {Yool}},\ }\href@noop {} {\bibfield
  {journal} {\bibinfo  {journal} {Biophys. Rep.}\ }\textbf {\bibinfo {volume}
  {3}},\ \bibinfo {pages} {100100} (\bibinfo {year} {2023})}\BibitemShut
  {NoStop}%
\bibitem [{\citenamefont {Murata}\ \emph {et~al.}(2000)\citenamefont {Murata},
  \citenamefont {Mitsuoka}, \citenamefont {Hirai}, \citenamefont {Walz},
  \citenamefont {Agre}, \citenamefont {Heymann}, \citenamefont {Engel},\ and\
  \citenamefont
  {Fujiyoshi}}]{MurataMitsuokaHiraiWalzAgreHeymannEngelFujiyoshi2000}%
  \BibitemOpen
  \bibfield  {author} {\bibinfo {author} {\bibfnamefont {K.}~\bibnamefont
  {Murata}}, \bibinfo {author} {\bibfnamefont {K.}~\bibnamefont {Mitsuoka}},
  \bibinfo {author} {\bibfnamefont {T.}~\bibnamefont {Hirai}}, \bibinfo
  {author} {\bibfnamefont {T.}~\bibnamefont {Walz}}, \bibinfo {author}
  {\bibfnamefont {P.}~\bibnamefont {Agre}}, \bibinfo {author} {\bibfnamefont
  {J.~B.}\ \bibnamefont {Heymann}}, \bibinfo {author} {\bibfnamefont
  {A.}~\bibnamefont {Engel}}, \ and\ \bibinfo {author} {\bibfnamefont
  {Y.}~\bibnamefont {Fujiyoshi}},\ }\href@noop {} {\bibfield  {journal}
  {\bibinfo  {journal} {Nature}\ }\textbf {\bibinfo {volume} {407}},\ \bibinfo
  {pages} {599} (\bibinfo {year} {2000})}\BibitemShut {NoStop}%
\bibitem [{\citenamefont {Sui}\ \emph {et~al.}(2001)\citenamefont {Sui},
  \citenamefont {Han}, \citenamefont {Lee}, \citenamefont {Walian},\ and\
  \citenamefont {Jap}}]{SuiHanLeeWalianJap2001}%
  \BibitemOpen
  \bibfield  {author} {\bibinfo {author} {\bibfnamefont {H.}~\bibnamefont
  {Sui}}, \bibinfo {author} {\bibfnamefont {B.~G.}\ \bibnamefont {Han}},
  \bibinfo {author} {\bibfnamefont {J.~K.}\ \bibnamefont {Lee}}, \bibinfo
  {author} {\bibfnamefont {P.}~\bibnamefont {Walian}}, \ and\ \bibinfo {author}
  {\bibfnamefont {B.~K.}\ \bibnamefont {Jap}},\ }\href@noop {} {\bibfield
  {journal} {\bibinfo  {journal} {Nature}\ }\textbf {\bibinfo {volume} {414}},\
  \bibinfo {pages} {872} (\bibinfo {year} {2001})}\BibitemShut {NoStop}%
\bibitem [{\citenamefont {Eriksson}\ \emph {et~al.}(2013)\citenamefont
  {Eriksson}, \citenamefont {Fischer}, \citenamefont {Friemann}, \citenamefont
  {Enkavi}, \citenamefont {Tajkhorshid},\ and\ \citenamefont
  {Neutze}}]{ErikssonFischerFriemannEnkaviTajkhorshidNeutze2013}%
  \BibitemOpen
  \bibfield  {author} {\bibinfo {author} {\bibfnamefont {U.~K.}\ \bibnamefont
  {Eriksson}}, \bibinfo {author} {\bibfnamefont {G.}~\bibnamefont {Fischer}},
  \bibinfo {author} {\bibfnamefont {R.}~\bibnamefont {Friemann}}, \bibinfo
  {author} {\bibfnamefont {G.}~\bibnamefont {Enkavi}}, \bibinfo {author}
  {\bibfnamefont {E.}~\bibnamefont {Tajkhorshid}}, \ and\ \bibinfo {author}
  {\bibfnamefont {R.}~\bibnamefont {Neutze}},\ }\href@noop {} {\bibfield
  {journal} {\bibinfo  {journal} {Science}\ }\textbf {\bibinfo {volume}
  {340}},\ \bibinfo {pages} {1346} (\bibinfo {year} {2013})}\BibitemShut
  {NoStop}%
\bibitem [{\citenamefont {Jensen}\ \emph {et~al.}(2003)\citenamefont {Jensen},
  \citenamefont {Tajkhorshid},\ and\ \citenamefont
  {Schulten}}]{JensenTajkhorshidSchulten2003}%
  \BibitemOpen
  \bibfield  {author} {\bibinfo {author} {\bibfnamefont {M.~{\O}.}\
  \bibnamefont {Jensen}}, \bibinfo {author} {\bibfnamefont {E.}~\bibnamefont
  {Tajkhorshid}}, \ and\ \bibinfo {author} {\bibfnamefont {K.}~\bibnamefont
  {Schulten}},\ }\href@noop {} {\bibfield  {journal} {\bibinfo  {journal}
  {Biophys. J.}\ }\textbf {\bibinfo {volume} {85}},\ \bibinfo {pages} {2884}
  (\bibinfo {year} {2003})}\BibitemShut {NoStop}%
\bibitem [{\citenamefont {Hub}\ \emph {et~al.}(2010)\citenamefont {Hub},
  \citenamefont {Aponte-Santamar{\'\i}a}, \citenamefont {Grubm{\"u}ller},\ and\
  \citenamefont {de~Groot}}]{HubAponte-SantamariaGrubmullerGroot2010}%
  \BibitemOpen
  \bibfield  {author} {\bibinfo {author} {\bibfnamefont {J.~S.}\ \bibnamefont
  {Hub}}, \bibinfo {author} {\bibfnamefont {C.}~\bibnamefont
  {Aponte-Santamar{\'\i}a}}, \bibinfo {author} {\bibfnamefont {H.}~\bibnamefont
  {Grubm{\"u}ller}}, \ and\ \bibinfo {author} {\bibfnamefont {B.~L.}\
  \bibnamefont {de~Groot}},\ }\href@noop {} {\bibfield  {journal} {\bibinfo
  {journal} {Biophys. J.}\ }\textbf {\bibinfo {volume} {99}},\ \bibinfo {pages}
  {L97} (\bibinfo {year} {2010})}\BibitemShut {NoStop}%
\bibitem [{\citenamefont {Yamamoto}\ \emph {et~al.}(2014)\citenamefont
  {Yamamoto}, \citenamefont {Akimoto}, \citenamefont {Hirano}, \citenamefont
  {Yasui},\ and\ \citenamefont
  {Yasuoka}}]{YamamotoAkimotoHiranoYasuiYasuoka2014}%
  \BibitemOpen
  \bibfield  {author} {\bibinfo {author} {\bibfnamefont {E.}~\bibnamefont
  {Yamamoto}}, \bibinfo {author} {\bibfnamefont {T.}~\bibnamefont {Akimoto}},
  \bibinfo {author} {\bibfnamefont {Y.}~\bibnamefont {Hirano}}, \bibinfo
  {author} {\bibfnamefont {M.}~\bibnamefont {Yasui}}, \ and\ \bibinfo {author}
  {\bibfnamefont {K.}~\bibnamefont {Yasuoka}},\ }\href {\doibase
  10.1103/PhysRevE.89.022718} {\bibfield  {journal} {\bibinfo  {journal} {Phys.
  Rev. E}\ }\textbf {\bibinfo {volume} {89}},\ \bibinfo {pages} {022718}
  (\bibinfo {year} {2014})}\BibitemShut {NoStop}%
\bibitem [{\citenamefont {Ma}\ \emph {et~al.}(1996)\citenamefont {Ma},
  \citenamefont {Yang}, \citenamefont {Kuo},\ and\ \citenamefont
  {Verkman}}]{MaYangKuoVerkman1996}%
  \BibitemOpen
  \bibfield  {author} {\bibinfo {author} {\bibfnamefont {T.}~\bibnamefont
  {Ma}}, \bibinfo {author} {\bibfnamefont {B.}~\bibnamefont {Yang}}, \bibinfo
  {author} {\bibfnamefont {W.-L.}\ \bibnamefont {Kuo}}, \ and\ \bibinfo
  {author} {\bibfnamefont {A.~S.}\ \bibnamefont {Verkman}},\ }\href@noop {}
  {\bibfield  {journal} {\bibinfo  {journal} {Genomics}\ }\textbf {\bibinfo
  {volume} {35}},\ \bibinfo {pages} {543} (\bibinfo {year} {1996})}\BibitemShut
  {NoStop}%
\bibitem [{\citenamefont {Yasui}\ \emph
  {et~al.}(1999{\natexlab{b}})\citenamefont {Yasui}, \citenamefont {Kwon},
  \citenamefont {Knepper}, \citenamefont {Nielsen},\ and\ \citenamefont
  {Agre}}]{YasuiKwonKnepperNielsenAgre1999}%
  \BibitemOpen
  \bibfield  {author} {\bibinfo {author} {\bibfnamefont {M.}~\bibnamefont
  {Yasui}}, \bibinfo {author} {\bibfnamefont {T.-H.}\ \bibnamefont {Kwon}},
  \bibinfo {author} {\bibfnamefont {M.~A.}\ \bibnamefont {Knepper}}, \bibinfo
  {author} {\bibfnamefont {S.}~\bibnamefont {Nielsen}}, \ and\ \bibinfo
  {author} {\bibfnamefont {P.}~\bibnamefont {Agre}},\ }\href@noop {} {\bibfield
   {journal} {\bibinfo  {journal} {Proc. Natl. Acad. Sci. USA}\ }\textbf
  {\bibinfo {volume} {96}},\ \bibinfo {pages} {5808} (\bibinfo {year}
  {1999}{\natexlab{b}})}\BibitemShut {NoStop}%
\bibitem [{\citenamefont {Kim}\ \emph {et~al.}(2011)\citenamefont {Kim},
  \citenamefont {Oh}, \citenamefont {Lee}, \citenamefont {Ahn}, \citenamefont
  {Kim},\ and\ \citenamefont {Park}}]{KimOhLeeAhnKimPark2011}%
  \BibitemOpen
  \bibfield  {author} {\bibinfo {author} {\bibfnamefont {S.-O.}\ \bibnamefont
  {Kim}}, \bibinfo {author} {\bibfnamefont {K.~J.}\ \bibnamefont {Oh}},
  \bibinfo {author} {\bibfnamefont {H.~S.}\ \bibnamefont {Lee}}, \bibinfo
  {author} {\bibfnamefont {K.}~\bibnamefont {Ahn}}, \bibinfo {author}
  {\bibfnamefont {S.~W.}\ \bibnamefont {Kim}}, \ and\ \bibinfo {author}
  {\bibfnamefont {K.}~\bibnamefont {Park}},\ }\href@noop {} {\bibfield
  {journal} {\bibinfo  {journal} {J. Sex. Med.}\ }\textbf {\bibinfo {volume}
  {8}},\ \bibinfo {pages} {1925} (\bibinfo {year} {2011})}\BibitemShut
  {NoStop}%
\bibitem [{\citenamefont {Ma}\ \emph {et~al.}(2016)\citenamefont {Ma},
  \citenamefont {Zhou}, \citenamefont {Yang}, \citenamefont {Ding},
  \citenamefont {Zhu},\ and\ \citenamefont {Chen}}]{MaZhouYangDingZhuChen2016}%
  \BibitemOpen
  \bibfield  {author} {\bibinfo {author} {\bibfnamefont {J.}~\bibnamefont
  {Ma}}, \bibinfo {author} {\bibfnamefont {C.}~\bibnamefont {Zhou}}, \bibinfo
  {author} {\bibfnamefont {J.}~\bibnamefont {Yang}}, \bibinfo {author}
  {\bibfnamefont {X.}~\bibnamefont {Ding}}, \bibinfo {author} {\bibfnamefont
  {Y.}~\bibnamefont {Zhu}}, \ and\ \bibinfo {author} {\bibfnamefont
  {X.}~\bibnamefont {Chen}},\ }\href@noop {} {\bibfield  {journal} {\bibinfo
  {journal} {J. Mol. Histol.}\ }\textbf {\bibinfo {volume} {47}},\ \bibinfo
  {pages} {129} (\bibinfo {year} {2016})}\BibitemShut {NoStop}%
\bibitem [{\citenamefont {Michalek}(2016)}]{Michalek2016}%
  \BibitemOpen
  \bibfield  {author} {\bibinfo {author} {\bibfnamefont {K.}~\bibnamefont
  {Michalek}},\ }\href@noop {} {\bibfield  {journal} {\bibinfo  {journal} {J.
  Physiol. Pharmacol.}\ }\textbf {\bibinfo {volume} {67}},\ \bibinfo {pages}
  {185} (\bibinfo {year} {2016})}\BibitemShut {NoStop}%
\bibitem [{\citenamefont {Ribeiro}\ \emph {et~al.}(2021)\citenamefont
  {Ribeiro}, \citenamefont {Alves}, \citenamefont {Yeste}, \citenamefont {Cho},
  \citenamefont {Calamita},\ and\ \citenamefont
  {Oliveira}}]{RibeiroAlvesYesteChoCalamitaOliveira2021}%
  \BibitemOpen
  \bibfield  {author} {\bibinfo {author} {\bibfnamefont {J.~C.}\ \bibnamefont
  {Ribeiro}}, \bibinfo {author} {\bibfnamefont {M.~G.}\ \bibnamefont {Alves}},
  \bibinfo {author} {\bibfnamefont {M.}~\bibnamefont {Yeste}}, \bibinfo
  {author} {\bibfnamefont {Y.~S.}\ \bibnamefont {Cho}}, \bibinfo {author}
  {\bibfnamefont {G.}~\bibnamefont {Calamita}}, \ and\ \bibinfo {author}
  {\bibfnamefont {P.~F.}\ \bibnamefont {Oliveira}},\ }\href@noop {} {\bibfield
  {journal} {\bibinfo  {journal} {Biochim. Biophys. Acta Mol. Basis. Dis.}\
  }\textbf {\bibinfo {volume} {1867}},\ \bibinfo {pages} {166039} (\bibinfo
  {year} {2021})}\BibitemShut {NoStop}%
\bibitem [{\citenamefont {Ikeda}\ \emph {et~al.}(2002)\citenamefont {Ikeda},
  \citenamefont {Beitz}, \citenamefont {Kozono}, \citenamefont {Guggino},
  \citenamefont {Agre},\ and\ \citenamefont
  {Yasui}}]{IkedaBeitzKozonoGugginoAgreYasui2002}%
  \BibitemOpen
  \bibfield  {author} {\bibinfo {author} {\bibfnamefont {M.}~\bibnamefont
  {Ikeda}}, \bibinfo {author} {\bibfnamefont {E.}~\bibnamefont {Beitz}},
  \bibinfo {author} {\bibfnamefont {D.}~\bibnamefont {Kozono}}, \bibinfo
  {author} {\bibfnamefont {W.~B.}\ \bibnamefont {Guggino}}, \bibinfo {author}
  {\bibfnamefont {P.}~\bibnamefont {Agre}}, \ and\ \bibinfo {author}
  {\bibfnamefont {M.}~\bibnamefont {Yasui}},\ }\href@noop {} {\bibfield
  {journal} {\bibinfo  {journal} {J. Biol. Chem.}\ }\textbf {\bibinfo {volume}
  {277}},\ \bibinfo {pages} {39873} (\bibinfo {year} {2002})}\BibitemShut
  {NoStop}%
\bibitem [{\citenamefont {Liu}\ \emph {et~al.}(2005)\citenamefont {Liu},
  \citenamefont {Kozono}, \citenamefont {Kato}, \citenamefont {Agre},
  \citenamefont {Hazama},\ and\ \citenamefont
  {Yasui}}]{LiuKozonoKatoAgreHazamaYasui2005}%
  \BibitemOpen
  \bibfield  {author} {\bibinfo {author} {\bibfnamefont {K.}~\bibnamefont
  {Liu}}, \bibinfo {author} {\bibfnamefont {D.}~\bibnamefont {Kozono}},
  \bibinfo {author} {\bibfnamefont {Y.}~\bibnamefont {Kato}}, \bibinfo {author}
  {\bibfnamefont {P.}~\bibnamefont {Agre}}, \bibinfo {author} {\bibfnamefont
  {A.}~\bibnamefont {Hazama}}, \ and\ \bibinfo {author} {\bibfnamefont
  {M.}~\bibnamefont {Yasui}},\ }\href@noop {} {\bibfield  {journal} {\bibinfo
  {journal} {Proc. Natl. Acad. Sci. USA}\ }\textbf {\bibinfo {volume} {102}},\
  \bibinfo {pages} {2192} (\bibinfo {year} {2005})}\BibitemShut {NoStop}%
\bibitem [{\citenamefont {Frick}\ \emph {et~al.}(2014)\citenamefont {Frick},
  \citenamefont {Eriksson}, \citenamefont {de~Mattia}, \citenamefont
  {{\"O}berg}, \citenamefont {Hedfalk}, \citenamefont {Neutze}, \citenamefont
  {Willem}, \citenamefont {Deen},\ and\ \citenamefont
  {T{\"o}rnroth-Horsefield}}]{FrickErikssonMattiaObergHedfalkNeutzeWillemDeenTornroth-Horsefield2014}%
  \BibitemOpen
  \bibfield  {author} {\bibinfo {author} {\bibfnamefont {A.}~\bibnamefont
  {Frick}}, \bibinfo {author} {\bibfnamefont {U.~K.}\ \bibnamefont {Eriksson}},
  \bibinfo {author} {\bibfnamefont {F.}~\bibnamefont {de~Mattia}}, \bibinfo
  {author} {\bibfnamefont {F.}~\bibnamefont {{\"O}berg}}, \bibinfo {author}
  {\bibfnamefont {K.}~\bibnamefont {Hedfalk}}, \bibinfo {author} {\bibfnamefont
  {R.}~\bibnamefont {Neutze}}, \bibinfo {author} {\bibfnamefont
  {J.}~\bibnamefont {Willem}}, \bibinfo {author} {\bibfnamefont {P.~M.~T.}\
  \bibnamefont {Deen}}, \ and\ \bibinfo {author} {\bibfnamefont
  {S.}~\bibnamefont {T{\"o}rnroth-Horsefield}},\ }\href@noop {} {\bibfield
  {journal} {\bibinfo  {journal} {Proc. Natl. Acad. Sci. USA}\ }\textbf
  {\bibinfo {volume} {111}},\ \bibinfo {pages} {6305} (\bibinfo {year}
  {2014})}\BibitemShut {NoStop}%
\bibitem [{\citenamefont {Horsefield}\ \emph {et~al.}(2008)\citenamefont
  {Horsefield}, \citenamefont {Nord{\'e}n}, \citenamefont {Fellert},
  \citenamefont {Backmark}, \citenamefont {T{\"o}rnroth-Horsefield},
  \citenamefont {van Scheltinga}, \citenamefont {Kvassman}, \citenamefont
  {Kjellbom}, \citenamefont {Johanson},\ and\ \citenamefont
  {Neutze}}]{HorsefieldNordenFellertBackmarkTornroth-HorsefieldScheltingaKvassmanKjellbomJohansonNeutze2008}%
  \BibitemOpen
  \bibfield  {author} {\bibinfo {author} {\bibfnamefont {R.}~\bibnamefont
  {Horsefield}}, \bibinfo {author} {\bibfnamefont {K.}~\bibnamefont
  {Nord{\'e}n}}, \bibinfo {author} {\bibfnamefont {M.}~\bibnamefont {Fellert}},
  \bibinfo {author} {\bibfnamefont {A.}~\bibnamefont {Backmark}}, \bibinfo
  {author} {\bibfnamefont {S.}~\bibnamefont {T{\"o}rnroth-Horsefield}},
  \bibinfo {author} {\bibfnamefont {A.~C.~T.}\ \bibnamefont {van Scheltinga}},
  \bibinfo {author} {\bibfnamefont {J.}~\bibnamefont {Kvassman}}, \bibinfo
  {author} {\bibfnamefont {P.}~\bibnamefont {Kjellbom}}, \bibinfo {author}
  {\bibfnamefont {U.}~\bibnamefont {Johanson}}, \ and\ \bibinfo {author}
  {\bibfnamefont {R.}~\bibnamefont {Neutze}},\ }\href@noop {} {\bibfield
  {journal} {\bibinfo  {journal} {Proc. Natl. Acad. Sci. USA}\ }\textbf
  {\bibinfo {volume} {105}},\ \bibinfo {pages} {13327} (\bibinfo {year}
  {2008})}\BibitemShut {NoStop}%
\bibitem [{\citenamefont {Nyblom}\ \emph {et~al.}(2009)\citenamefont {Nyblom},
  \citenamefont {Frick}, \citenamefont {Wang}, \citenamefont {Ekvall},
  \citenamefont {Hallgren}, \citenamefont {Hedfalk}, \citenamefont {Neutze},
  \citenamefont {Tajkhorshid},\ and\ \citenamefont
  {T{\"o}rnroth-Horsefield}}]{NyblomFrickWangEkvallHallgrenHedfalkNeutzeTajkhorshidTornroth-Horsefield2009}%
  \BibitemOpen
  \bibfield  {author} {\bibinfo {author} {\bibfnamefont {M.}~\bibnamefont
  {Nyblom}}, \bibinfo {author} {\bibfnamefont {A.}~\bibnamefont {Frick}},
  \bibinfo {author} {\bibfnamefont {Y.}~\bibnamefont {Wang}}, \bibinfo {author}
  {\bibfnamefont {M.}~\bibnamefont {Ekvall}}, \bibinfo {author} {\bibfnamefont
  {K.}~\bibnamefont {Hallgren}}, \bibinfo {author} {\bibfnamefont
  {K.}~\bibnamefont {Hedfalk}}, \bibinfo {author} {\bibfnamefont
  {R.}~\bibnamefont {Neutze}}, \bibinfo {author} {\bibfnamefont
  {E.}~\bibnamefont {Tajkhorshid}}, \ and\ \bibinfo {author} {\bibfnamefont
  {S.}~\bibnamefont {T{\"o}rnroth-Horsefield}},\ }\href@noop {} {\bibfield
  {journal} {\bibinfo  {journal} {J. Mol. Biol.}\ }\textbf {\bibinfo {volume}
  {387}},\ \bibinfo {pages} {653} (\bibinfo {year} {2009})}\BibitemShut
  {NoStop}%
\bibitem [{\citenamefont {Kirscht}\ \emph {et~al.}(2016)\citenamefont
  {Kirscht}, \citenamefont {Kaptan}, \citenamefont {Bienert}, \citenamefont
  {Chaumont}, \citenamefont {Nissen}, \citenamefont {de~Groot}, \citenamefont
  {Kjellbom}, \citenamefont {Gourdon},\ and\ \citenamefont
  {Johanson}}]{KirschtKaptanBienertChaumontNissenGrootKjellbomGourdonJohanson2016}%
  \BibitemOpen
  \bibfield  {author} {\bibinfo {author} {\bibfnamefont {A.}~\bibnamefont
  {Kirscht}}, \bibinfo {author} {\bibfnamefont {S.~S.}\ \bibnamefont {Kaptan}},
  \bibinfo {author} {\bibfnamefont {G.~P.}\ \bibnamefont {Bienert}}, \bibinfo
  {author} {\bibfnamefont {F.}~\bibnamefont {Chaumont}}, \bibinfo {author}
  {\bibfnamefont {P.}~\bibnamefont {Nissen}}, \bibinfo {author} {\bibfnamefont
  {B.~L.}\ \bibnamefont {de~Groot}}, \bibinfo {author} {\bibfnamefont
  {P.}~\bibnamefont {Kjellbom}}, \bibinfo {author} {\bibfnamefont
  {P.}~\bibnamefont {Gourdon}}, \ and\ \bibinfo {author} {\bibfnamefont
  {U.}~\bibnamefont {Johanson}},\ }\href@noop {} {\bibfield  {journal}
  {\bibinfo  {journal} {PLoS Biol.}\ }\textbf {\bibinfo {volume} {14}},\
  \bibinfo {pages} {e1002411} (\bibinfo {year} {2016})}\BibitemShut {NoStop}%
\bibitem [{\citenamefont {Kutzner}\ \emph {et~al.}(2011)\citenamefont
  {Kutzner}, \citenamefont {Grubm{\\"u}ller}, \citenamefont {de~Groot},\ and\
  \citenamefont {Zachariae}}]{KutznerGrubmullerGrootZachariae2011}%
  \BibitemOpen
  \bibfield  {author} {\bibinfo {author} {\bibfnamefont {C.}~\bibnamefont
  {Kutzner}}, \bibinfo {author} {\bibfnamefont {H.}~\bibnamefont
  {Grubm{\\"u}ller}}, \bibinfo {author} {\bibfnamefont {B.}~\bibnamefont
  {de~Groot}}, \ and\ \bibinfo {author} {\bibfnamefont {U.}~\bibnamefont
  {Zachariae}},\ }\href@noop {} {\bibfield  {journal} {\bibinfo  {journal}
  {Biophys. J.}\ }\textbf {\bibinfo {volume} {101}},\ \bibinfo {pages} {809}
  (\bibinfo {year} {2011})}\BibitemShut {NoStop}%
\bibitem [{\citenamefont {Kutzner}\ \emph {et~al.}(2016)\citenamefont
  {Kutzner}, \citenamefont {K{\"o}pfer}, \citenamefont {Machtens},
  \citenamefont {de~Groot}, \citenamefont {Song},\ and\ \citenamefont
  {Zachariae}}]{KutznerKoepferMachtensGrootSongZachariae2016}%
  \BibitemOpen
  \bibfield  {author} {\bibinfo {author} {\bibfnamefont {C.}~\bibnamefont
  {Kutzner}}, \bibinfo {author} {\bibfnamefont {D.~A.}\ \bibnamefont
  {K{\"o}pfer}}, \bibinfo {author} {\bibfnamefont {J.-P.}\ \bibnamefont
  {Machtens}}, \bibinfo {author} {\bibfnamefont {B.~L.}\ \bibnamefont
  {de~Groot}}, \bibinfo {author} {\bibfnamefont {C.}~\bibnamefont {Song}}, \
  and\ \bibinfo {author} {\bibfnamefont {U.}~\bibnamefont {Zachariae}},\
  }\href@noop {} {\bibfield  {journal} {\bibinfo  {journal} {Biochim. Biophys.
  Acta}\ }\textbf {\bibinfo {volume} {1858}},\ \bibinfo {pages} {1741}
  (\bibinfo {year} {2016})}\BibitemShut {NoStop}%
\bibitem [{\citenamefont {Machtens}\ \emph {et~al.}(2015)\citenamefont
  {Machtens}, \citenamefont {Kortzak}, \citenamefont {Lansche}, \citenamefont
  {Leinenweber}, \citenamefont {Kilian}, \citenamefont {Begemann},
  \citenamefont {Zachariae}, \citenamefont {Ewers}, \citenamefont {de~Groot},
  \citenamefont {Briones},\ and\ \citenamefont
  {Fahlke}}]{MachtensKortzakLanscheLeinenweberKilianBegemannZachariaeEwersGrootBrionesFahlke2015}%
  \BibitemOpen
  \bibfield  {author} {\bibinfo {author} {\bibfnamefont {J.-P.}\ \bibnamefont
  {Machtens}}, \bibinfo {author} {\bibfnamefont {D.}~\bibnamefont {Kortzak}},
  \bibinfo {author} {\bibfnamefont {C.}~\bibnamefont {Lansche}}, \bibinfo
  {author} {\bibfnamefont {A.}~\bibnamefont {Leinenweber}}, \bibinfo {author}
  {\bibfnamefont {P.}~\bibnamefont {Kilian}}, \bibinfo {author} {\bibfnamefont
  {B.}~\bibnamefont {Begemann}}, \bibinfo {author} {\bibfnamefont
  {U.}~\bibnamefont {Zachariae}}, \bibinfo {author} {\bibfnamefont
  {D.}~\bibnamefont {Ewers}}, \bibinfo {author} {\bibfnamefont {B.~L.}\
  \bibnamefont {de~Groot}}, \bibinfo {author} {\bibfnamefont {R.}~\bibnamefont
  {Briones}}, \ and\ \bibinfo {author} {\bibfnamefont {C.}~\bibnamefont
  {Fahlke}},\ }\href@noop {} {\bibfield  {journal} {\bibinfo  {journal} {Cell}\
  }\textbf {\bibinfo {volume} {160}},\ \bibinfo {pages} {542} (\bibinfo {year}
  {2015})}\BibitemShut {NoStop}%
\bibitem [{\citenamefont {Klesse}\ \emph
  {et~al.}(2020{\natexlab{a}})\citenamefont {Klesse}, \citenamefont {Tucker},\
  and\ \citenamefont {Sansom}}]{KlesseTuckerSansom2020}%
  \BibitemOpen
  \bibfield  {author} {\bibinfo {author} {\bibfnamefont {G.}~\bibnamefont
  {Klesse}}, \bibinfo {author} {\bibfnamefont {S.~J.}\ \bibnamefont {Tucker}},
  \ and\ \bibinfo {author} {\bibfnamefont {M.~S.~P.}\ \bibnamefont {Sansom}},\
  }\href@noop {} {\bibfield  {journal} {\bibinfo  {journal} {ACS Nano}\
  }\textbf {\bibinfo {volume} {14}},\ \bibinfo {pages} {10480} (\bibinfo {year}
  {2020}{\natexlab{a}})}\BibitemShut {NoStop}%
\bibitem [{\citenamefont {Lynch}\ \emph {et~al.}(2020)\citenamefont {Lynch},
  \citenamefont {Rao},\ and\ \citenamefont {Sansom}}]{LynchRaoSansom2020}%
  \BibitemOpen
  \bibfield  {author} {\bibinfo {author} {\bibfnamefont {C.~I.}\ \bibnamefont
  {Lynch}}, \bibinfo {author} {\bibfnamefont {S.}~\bibnamefont {Rao}}, \ and\
  \bibinfo {author} {\bibfnamefont {M.~S.~P.}\ \bibnamefont {Sansom}},\
  }\href@noop {} {\bibfield  {journal} {\bibinfo  {journal} {Chem. Rev.}\
  }\textbf {\bibinfo {volume} {120}},\ \bibinfo {pages} {10298} (\bibinfo
  {year} {2020})}\BibitemShut {NoStop}%
\bibitem [{\citenamefont {Klesse}\ \emph
  {et~al.}(2020{\natexlab{b}})\citenamefont {Klesse}, \citenamefont {Rao},
  \citenamefont {Tucker},\ and\ \citenamefont
  {Sansom}}]{KlesseRaoTuckerSansom2020}%
  \BibitemOpen
  \bibfield  {author} {\bibinfo {author} {\bibfnamefont {G.}~\bibnamefont
  {Klesse}}, \bibinfo {author} {\bibfnamefont {S.}~\bibnamefont {Rao}},
  \bibinfo {author} {\bibfnamefont {S.~J.}\ \bibnamefont {Tucker}}, \ and\
  \bibinfo {author} {\bibfnamefont {M.~S.~P.}\ \bibnamefont {Sansom}},\
  }\href@noop {} {\bibfield  {journal} {\bibinfo  {journal} {J. Am. Chem.
  Soc.}\ }\textbf {\bibinfo {volume} {142}},\ \bibinfo {pages} {9415} (\bibinfo
  {year} {2020}{\natexlab{b}})}\BibitemShut {NoStop}%
\bibitem [{\citenamefont {Jojoa-Cruz}\ \emph {et~al.}(2022)\citenamefont
  {Jojoa-Cruz}, \citenamefont {Saotome}, \citenamefont {Tsui}, \citenamefont
  {Lee}, \citenamefont {Sansom}, \citenamefont {Murthy}, \citenamefont
  {Patapoutian},\ and\ \citenamefont
  {Ward}}]{JojoaCruzSaotomeTsuiLeeSansomMurthyPatapoutianWard2022}%
  \BibitemOpen
  \bibfield  {author} {\bibinfo {author} {\bibfnamefont {S.}~\bibnamefont
  {Jojoa-Cruz}}, \bibinfo {author} {\bibfnamefont {K.}~\bibnamefont {Saotome}},
  \bibinfo {author} {\bibfnamefont {C.~C.~A.}\ \bibnamefont {Tsui}}, \bibinfo
  {author} {\bibfnamefont {W.-H.}\ \bibnamefont {Lee}}, \bibinfo {author}
  {\bibfnamefont {M.~S.~P.}\ \bibnamefont {Sansom}}, \bibinfo {author}
  {\bibfnamefont {S.~E.}\ \bibnamefont {Murthy}}, \bibinfo {author}
  {\bibfnamefont {A.}~\bibnamefont {Patapoutian}}, \ and\ \bibinfo {author}
  {\bibfnamefont {A.~B.}\ \bibnamefont {Ward}},\ }\href@noop {} {\bibfield
  {journal} {\bibinfo  {journal} {Nat. Commun.}\ }\textbf {\bibinfo {volume}
  {13}},\ \bibinfo {pages} {850} (\bibinfo {year} {2022})}\BibitemShut
  {NoStop}%
\bibitem [{\citenamefont {Phan}\ \emph {et~al.}(2023)\citenamefont {Phan},
  \citenamefont {Chamorro}, \citenamefont {Martinez-Seara}, \citenamefont
  {Crain}, \citenamefont {Sansom},\ and\ \citenamefont
  {Tucker}}]{PhanChamorroMartinezSearaCrainSansomTucker2023}%
  \BibitemOpen
  \bibfield  {author} {\bibinfo {author} {\bibfnamefont {L.~X.}\ \bibnamefont
  {Phan}}, \bibinfo {author} {\bibfnamefont {V.~C.}\ \bibnamefont {Chamorro}},
  \bibinfo {author} {\bibfnamefont {H.}~\bibnamefont {Martinez-Seara}},
  \bibinfo {author} {\bibfnamefont {J.}~\bibnamefont {Crain}}, \bibinfo
  {author} {\bibfnamefont {M.~S.~P.}\ \bibnamefont {Sansom}}, \ and\ \bibinfo
  {author} {\bibfnamefont {S.~J.}\ \bibnamefont {Tucker}},\ }\href@noop {}
  {\bibfield  {journal} {\bibinfo  {journal} {Biophys. J.}\ }\textbf {\bibinfo
  {volume} {122}},\ \bibinfo {pages} {1548} (\bibinfo {year}
  {2023})}\BibitemShut {NoStop}%
\bibitem [{\citenamefont {Fujiyoshi}\ \emph {et~al.}(2002)\citenamefont
  {Fujiyoshi}, \citenamefont {Mitsuoka}, \citenamefont {de~Groot},
  \citenamefont {Philippsen}, \citenamefont {Grubmuller}, \citenamefont
  {Agre},\ and\ \citenamefont
  {Engel}}]{FujiyoshiMitsuokaGrootPhilippsenGrubmullerAgreEngel2002}%
  \BibitemOpen
  \bibfield  {author} {\bibinfo {author} {\bibfnamefont {Y.}~\bibnamefont
  {Fujiyoshi}}, \bibinfo {author} {\bibfnamefont {K.}~\bibnamefont {Mitsuoka}},
  \bibinfo {author} {\bibfnamefont {B.}~\bibnamefont {de~Groot}}, \bibinfo
  {author} {\bibfnamefont {A.}~\bibnamefont {Philippsen}}, \bibinfo {author}
  {\bibfnamefont {H.}~\bibnamefont {Grubmuller}}, \bibinfo {author}
  {\bibfnamefont {P.}~\bibnamefont {Agre}}, \ and\ \bibinfo {author}
  {\bibfnamefont {A.}~\bibnamefont {Engel}},\ }\href@noop {} {\bibfield
  {journal} {\bibinfo  {journal} {Curr. Opin. Struct. Biol.}\ }\textbf
  {\bibinfo {volume} {12}},\ \bibinfo {pages} {509} (\bibinfo {year}
  {2002})}\BibitemShut {NoStop}%
\bibitem [{\citenamefont {Maffeo}\ \emph {et~al.}(2012)\citenamefont {Maffeo},
  \citenamefont {Bhattacharya}, \citenamefont {Yoo}, \citenamefont {Wells},\
  and\ \citenamefont
  {Aksimentiev}}]{MaffeoBhattacharyaYooWellsAksimentiev2012}%
  \BibitemOpen
  \bibfield  {author} {\bibinfo {author} {\bibfnamefont {C.}~\bibnamefont
  {Maffeo}}, \bibinfo {author} {\bibfnamefont {S.}~\bibnamefont
  {Bhattacharya}}, \bibinfo {author} {\bibfnamefont {J.}~\bibnamefont {Yoo}},
  \bibinfo {author} {\bibfnamefont {D.}~\bibnamefont {Wells}}, \ and\ \bibinfo
  {author} {\bibfnamefont {A.}~\bibnamefont {Aksimentiev}},\ }\href@noop {}
  {\bibfield  {journal} {\bibinfo  {journal} {Chem. Rev.}\ }\textbf {\bibinfo
  {volume} {112}},\ \bibinfo {pages} {6250} (\bibinfo {year}
  {2012})}\BibitemShut {NoStop}%
\bibitem [{\citenamefont {Trick}\ \emph {et~al.}(2016)\citenamefont {Trick},
  \citenamefont {Chelvaniththilan}, \citenamefont {Klesse}, \citenamefont
  {Aryal}, \citenamefont {Wallace}, \citenamefont {Tucker},\ and\ \citenamefont
  {Sansom}}]{TrickChelvaniththilanKlesseAryalWallaceTuckerSansom2016}%
  \BibitemOpen
  \bibfield  {author} {\bibinfo {author} {\bibfnamefont {J.~L.}\ \bibnamefont
  {Trick}}, \bibinfo {author} {\bibfnamefont {S.}~\bibnamefont
  {Chelvaniththilan}}, \bibinfo {author} {\bibfnamefont {G.}~\bibnamefont
  {Klesse}}, \bibinfo {author} {\bibfnamefont {P.}~\bibnamefont {Aryal}},
  \bibinfo {author} {\bibfnamefont {E.~J.}\ \bibnamefont {Wallace}}, \bibinfo
  {author} {\bibfnamefont {S.~J.}\ \bibnamefont {Tucker}}, \ and\ \bibinfo
  {author} {\bibfnamefont {M.~S.}\ \bibnamefont {Sansom}},\ }\href@noop {}
  {\bibfield  {journal} {\bibinfo  {journal} {Structure}\ }\textbf {\bibinfo
  {volume} {24}},\ \bibinfo {pages} {2207} (\bibinfo {year}
  {2016})}\BibitemShut {NoStop}%
\bibitem [{\citenamefont {Lynch}\ \emph {et~al.}(2021)\citenamefont {Lynch},
  \citenamefont {Klesse}, \citenamefont {Rao}, \citenamefont {Tucker},\ and\
  \citenamefont {Sansom}}]{LynchKlesseRaoTuckerSansom2021}%
  \BibitemOpen
  \bibfield  {author} {\bibinfo {author} {\bibfnamefont {C.~I.}\ \bibnamefont
  {Lynch}}, \bibinfo {author} {\bibfnamefont {G.}~\bibnamefont {Klesse}},
  \bibinfo {author} {\bibfnamefont {S.}~\bibnamefont {Rao}}, \bibinfo {author}
  {\bibfnamefont {S.~J.}\ \bibnamefont {Tucker}}, \ and\ \bibinfo {author}
  {\bibfnamefont {M.~S.~P.}\ \bibnamefont {Sansom}},\ }\href@noop {} {\bibfield
   {journal} {\bibinfo  {journal} {ACS Nano}\ }\textbf {\bibinfo {volume}
  {15}},\ \bibinfo {pages} {19098} (\bibinfo {year} {2021})}\BibitemShut
  {NoStop}%
\bibitem [{\citenamefont {Rao}\ \emph {et~al.}(2019)\citenamefont {Rao},
  \citenamefont {Klesse}, \citenamefont {Stansfeld}, \citenamefont {Tucker},\
  and\ \citenamefont {Sansom}}]{RaoKlesseStansfeldTuckerSansom2019}%
  \BibitemOpen
  \bibfield  {author} {\bibinfo {author} {\bibfnamefont {S.}~\bibnamefont
  {Rao}}, \bibinfo {author} {\bibfnamefont {G.}~\bibnamefont {Klesse}},
  \bibinfo {author} {\bibfnamefont {P.~J.}\ \bibnamefont {Stansfeld}}, \bibinfo
  {author} {\bibfnamefont {S.~J.}\ \bibnamefont {Tucker}}, \ and\ \bibinfo
  {author} {\bibfnamefont {M.~S.~P.}\ \bibnamefont {Sansom}},\ }\href@noop {}
  {\bibfield  {journal} {\bibinfo  {journal} {Proc. Natl. Acad. Sci. USA}\
  }\textbf {\bibinfo {volume} {116}},\ \bibinfo {pages} {13989} (\bibinfo
  {year} {2019})}\BibitemShut {NoStop}%
\bibitem [{\citenamefont {Arai}\ \emph {et~al.}(2023)\citenamefont {Arai},
  \citenamefont {Yamamoto}, \citenamefont {Koishi}, \citenamefont {Hirano},
  \citenamefont {Yasuoka},\ and\ \citenamefont
  {Ebisuzaki}}]{AraiYamamotoKoishiHiranoYasuokaEbisuzaki2023}%
  \BibitemOpen
  \bibfield  {author} {\bibinfo {author} {\bibfnamefont {N.}~\bibnamefont
  {Arai}}, \bibinfo {author} {\bibfnamefont {E.}~\bibnamefont {Yamamoto}},
  \bibinfo {author} {\bibfnamefont {T.}~\bibnamefont {Koishi}}, \bibinfo
  {author} {\bibfnamefont {Y.}~\bibnamefont {Hirano}}, \bibinfo {author}
  {\bibfnamefont {K.}~\bibnamefont {Yasuoka}}, \ and\ \bibinfo {author}
  {\bibfnamefont {T.}~\bibnamefont {Ebisuzaki}},\ }\href@noop {} {\bibfield
  {journal} {\bibinfo  {journal} {Nanoscale Horiz.}\ }\textbf {\bibinfo
  {volume} {8}},\ \bibinfo {pages} {652} (\bibinfo {year} {2023})}\BibitemShut
  {NoStop}%
\bibitem [{\citenamefont {Joo}\ \emph {et~al.}(2016)\citenamefont {Joo},
  \citenamefont {Joung}, \citenamefont {Lee}, \citenamefont {Kim},
  \citenamefont {Cheng}, \citenamefont {Manavalan}, \citenamefont {Joung},
  \citenamefont {Heo}, \citenamefont {Lee}, \citenamefont {Nam}, \citenamefont
  {Lee}, \citenamefont {Lee},\ and\ \citenamefont
  {Lee}}]{JooJoungLeeKimChengManavalanJoungHeoLeeNamLeeLeeLee2016}%
  \BibitemOpen
  \bibfield  {author} {\bibinfo {author} {\bibfnamefont {K.}~\bibnamefont
  {Joo}}, \bibinfo {author} {\bibfnamefont {I.}~\bibnamefont {Joung}}, \bibinfo
  {author} {\bibfnamefont {S.~Y.}\ \bibnamefont {Lee}}, \bibinfo {author}
  {\bibfnamefont {J.~Y.}\ \bibnamefont {Kim}}, \bibinfo {author} {\bibfnamefont
  {Q.}~\bibnamefont {Cheng}}, \bibinfo {author} {\bibfnamefont
  {B.}~\bibnamefont {Manavalan}}, \bibinfo {author} {\bibfnamefont {J.~Y.}\
  \bibnamefont {Joung}}, \bibinfo {author} {\bibfnamefont {S.}~\bibnamefont
  {Heo}}, \bibinfo {author} {\bibfnamefont {J.}~\bibnamefont {Lee}}, \bibinfo
  {author} {\bibfnamefont {M.}~\bibnamefont {Nam}}, \bibinfo {author}
  {\bibfnamefont {I.-H.}\ \bibnamefont {Lee}}, \bibinfo {author} {\bibfnamefont
  {S.~J.}\ \bibnamefont {Lee}}, \ and\ \bibinfo {author} {\bibfnamefont
  {J.}~\bibnamefont {Lee}},\ }\href@noop {} {\bibfield  {journal} {\bibinfo
  {journal} {Proteins}\ }\textbf {\bibinfo {volume} {84}},\ \bibinfo {pages}
  {221} (\bibinfo {year} {2016})}\BibitemShut {NoStop}%
\bibitem [{\citenamefont {Joo}\ \emph {et~al.}(2008)\citenamefont {Joo},
  \citenamefont {Lee}, \citenamefont {Kim}, \citenamefont {Lee},\ and\
  \citenamefont {Lee}}]{JooLeeKimLeeLee2008}%
  \BibitemOpen
  \bibfield  {author} {\bibinfo {author} {\bibfnamefont {K.}~\bibnamefont
  {Joo}}, \bibinfo {author} {\bibfnamefont {J.}~\bibnamefont {Lee}}, \bibinfo
  {author} {\bibfnamefont {I.}~\bibnamefont {Kim}}, \bibinfo {author}
  {\bibfnamefont {S.~J.}\ \bibnamefont {Lee}}, \ and\ \bibinfo {author}
  {\bibfnamefont {J.}~\bibnamefont {Lee}},\ }\href@noop {} {\bibfield
  {journal} {\bibinfo  {journal} {Biophys. J.}\ }\textbf {\bibinfo {volume}
  {95}},\ \bibinfo {pages} {4813} (\bibinfo {year} {2008})}\BibitemShut
  {NoStop}%
\bibitem [{\citenamefont {Joo}\ \emph {et~al.}(2009)\citenamefont {Joo},
  \citenamefont {Lee}, \citenamefont {Seo}, \citenamefont {Lee}, \citenamefont
  {Kim},\ and\ \citenamefont {Lee}}]{JooLeeSeoLeeKimLee2009}%
  \BibitemOpen
  \bibfield  {author} {\bibinfo {author} {\bibfnamefont {K.}~\bibnamefont
  {Joo}}, \bibinfo {author} {\bibfnamefont {J.}~\bibnamefont {Lee}}, \bibinfo
  {author} {\bibfnamefont {J.-H.}\ \bibnamefont {Seo}}, \bibinfo {author}
  {\bibfnamefont {K.}~\bibnamefont {Lee}}, \bibinfo {author} {\bibfnamefont
  {B.-G.}\ \bibnamefont {Kim}}, \ and\ \bibinfo {author} {\bibfnamefont
  {J.}~\bibnamefont {Lee}},\ }\href@noop {} {\bibfield  {journal} {\bibinfo
  {journal} {Proteins}\ }\textbf {\bibinfo {volume} {75}},\ \bibinfo {pages}
  {1010} (\bibinfo {year} {2009})}\BibitemShut {NoStop}%
\bibitem [{\citenamefont {Lee}\ \emph {et~al.}(1997)\citenamefont {Lee},
  \citenamefont {Scheraga},\ and\ \citenamefont
  {Rackovsky}}]{LeeScheragaRackovsky1997}%
  \BibitemOpen
  \bibfield  {author} {\bibinfo {author} {\bibfnamefont {J.}~\bibnamefont
  {Lee}}, \bibinfo {author} {\bibfnamefont {H.~A.}\ \bibnamefont {Scheraga}}, \
  and\ \bibinfo {author} {\bibfnamefont {S.}~\bibnamefont {Rackovsky}},\
  }\href@noop {} {\bibfield  {journal} {\bibinfo  {journal} {J. Comput. Chem.}\
  }\textbf {\bibinfo {volume} {18}},\ \bibinfo {pages} {1222} (\bibinfo {year}
  {1997})}\BibitemShut {NoStop}%
\bibitem [{\citenamefont {Parrinello}\ and\ \citenamefont
  {Rahman}(1981)}]{ParrinelloRahman1981}%
  \BibitemOpen
  \bibfield  {author} {\bibinfo {author} {\bibfnamefont {M.}~\bibnamefont
  {Parrinello}}\ and\ \bibinfo {author} {\bibfnamefont {A.}~\bibnamefont
  {Rahman}},\ }\href@noop {} {\bibfield  {journal} {\bibinfo  {journal} {J.
  Appl. Phys.}\ }\textbf {\bibinfo {volume} {52}},\ \bibinfo {pages} {7182}
  (\bibinfo {year} {1981})}\BibitemShut {NoStop}%
\bibitem [{\citenamefont {Bussi}\ \emph {et~al.}(2009)\citenamefont {Bussi},
  \citenamefont {Zykova-Timan},\ and\ \citenamefont
  {Parrinello}}]{BussiZykova-TimanParrinello2009}%
  \BibitemOpen
  \bibfield  {author} {\bibinfo {author} {\bibfnamefont {G.}~\bibnamefont
  {Bussi}}, \bibinfo {author} {\bibfnamefont {T.}~\bibnamefont {Zykova-Timan}},
  \ and\ \bibinfo {author} {\bibfnamefont {M.}~\bibnamefont {Parrinello}},\
  }\href@noop {} {\bibfield  {journal} {\bibinfo  {journal} {J. Chem. Phys.}\
  }\textbf {\bibinfo {volume} {130}},\ \bibinfo {pages} {074101} (\bibinfo
  {year} {2009})}\BibitemShut {NoStop}%
\bibitem [{\citenamefont {Hess}\ \emph {et~al.}(1997)\citenamefont {Hess},
  \citenamefont {Bekker}, \citenamefont {Berendsen},\ and\ \citenamefont
  {Fraaije}}]{HessBekkerBerendsenFraaije1997}%
  \BibitemOpen
  \bibfield  {author} {\bibinfo {author} {\bibfnamefont {B.}~\bibnamefont
  {Hess}}, \bibinfo {author} {\bibfnamefont {H.}~\bibnamefont {Bekker}},
  \bibinfo {author} {\bibfnamefont {H.~J.~C.}\ \bibnamefont {Berendsen}}, \
  and\ \bibinfo {author} {\bibfnamefont {J.~G. E.~M.}\ \bibnamefont
  {Fraaije}},\ }\href@noop {} {\bibfield  {journal} {\bibinfo  {journal} {J.
  Comput. Chem.}\ }\textbf {\bibinfo {volume} {18}},\ \bibinfo {pages} {1463}
  (\bibinfo {year} {1997})}\BibitemShut {NoStop}%
\bibitem [{\citenamefont {Lindorff-Larsen}\ \emph {et~al.}(2010)\citenamefont
  {Lindorff-Larsen}, \citenamefont {Piana}, \citenamefont {Palmo},
  \citenamefont {Maragakis}, \citenamefont {Klepeis}, \citenamefont {Dror},\
  and\ \citenamefont
  {Shaw}}]{Lindorff-LarsenPianaPalmoMaragakisKlepeisDrorShaw2010}%
  \BibitemOpen
  \bibfield  {author} {\bibinfo {author} {\bibfnamefont {K.}~\bibnamefont
  {Lindorff-Larsen}}, \bibinfo {author} {\bibfnamefont {S.}~\bibnamefont
  {Piana}}, \bibinfo {author} {\bibfnamefont {K.}~\bibnamefont {Palmo}},
  \bibinfo {author} {\bibfnamefont {P.}~\bibnamefont {Maragakis}}, \bibinfo
  {author} {\bibfnamefont {J.~L.}\ \bibnamefont {Klepeis}}, \bibinfo {author}
  {\bibfnamefont {R.~O.}\ \bibnamefont {Dror}}, \ and\ \bibinfo {author}
  {\bibfnamefont {D.~E.}\ \bibnamefont {Shaw}},\ }\href@noop {} {\bibfield
  {journal} {\bibinfo  {journal} {Proteins}\ }\textbf {\bibinfo {volume}
  {78}},\ \bibinfo {pages} {1950} (\bibinfo {year} {2010})}\BibitemShut
  {NoStop}%
\bibitem [{\citenamefont {J{\"a}mbeck}\ and\ \citenamefont
  {Lyubartsev}(2012)}]{JambeckLyubartsev2012a}%
  \BibitemOpen
  \bibfield  {author} {\bibinfo {author} {\bibfnamefont {J.~P.~M.}\
  \bibnamefont {J{\"a}mbeck}}\ and\ \bibinfo {author} {\bibfnamefont {A.~P.}\
  \bibnamefont {Lyubartsev}},\ }\href@noop {} {\bibfield  {journal} {\bibinfo
  {journal} {J. Chem. Theory Comput.}\ }\textbf {\bibinfo {volume} {8}},\
  \bibinfo {pages} {2938} (\bibinfo {year} {2012})}\BibitemShut {NoStop}%
\bibitem [{\citenamefont {Jorgensen}\ \emph {et~al.}(1983)\citenamefont
  {Jorgensen}, \citenamefont {Chandrasekhar}, \citenamefont {Madura},
  \citenamefont {Impey},\ and\ \citenamefont
  {Klein}}]{JorgensenChandrasekharMaduraImpeyKlein1983}%
  \BibitemOpen
  \bibfield  {author} {\bibinfo {author} {\bibfnamefont {W.~L.}\ \bibnamefont
  {Jorgensen}}, \bibinfo {author} {\bibfnamefont {J.}~\bibnamefont
  {Chandrasekhar}}, \bibinfo {author} {\bibfnamefont {J.~D.}\ \bibnamefont
  {Madura}}, \bibinfo {author} {\bibfnamefont {R.~W.}\ \bibnamefont {Impey}}, \
  and\ \bibinfo {author} {\bibfnamefont {M.~L.}\ \bibnamefont {Klein}},\
  }\href@noop {} {\bibfield  {journal} {\bibinfo  {journal} {J. Chem. Phys.}\
  }\textbf {\bibinfo {volume} {79}},\ \bibinfo {pages} {926} (\bibinfo {year}
  {1983})}\BibitemShut {NoStop}%
\bibitem [{\citenamefont {Essmann}\ \emph {et~al.}(1995)\citenamefont
  {Essmann}, \citenamefont {Perera}, \citenamefont {Berkowitz}, \citenamefont
  {Darden}, \citenamefont {Lee},\ and\ \citenamefont
  {Pedersen}}]{EssmannPereraBerkowitzDardenLeePedersen1995}%
  \BibitemOpen
  \bibfield  {author} {\bibinfo {author} {\bibfnamefont {U.}~\bibnamefont
  {Essmann}}, \bibinfo {author} {\bibfnamefont {L.}~\bibnamefont {Perera}},
  \bibinfo {author} {\bibfnamefont {M.~L.}\ \bibnamefont {Berkowitz}}, \bibinfo
  {author} {\bibfnamefont {T.}~\bibnamefont {Darden}}, \bibinfo {author}
  {\bibfnamefont {H.}~\bibnamefont {Lee}}, \ and\ \bibinfo {author}
  {\bibfnamefont {L.~G.}\ \bibnamefont {Pedersen}},\ }\href@noop {} {\bibfield
  {journal} {\bibinfo  {journal} {J. Chem. Phys.}\ }\textbf {\bibinfo {volume}
  {103}},\ \bibinfo {pages} {8577} (\bibinfo {year} {1995})}\BibitemShut
  {NoStop}%
\end{thebibliography}

%

\setcounter{figure}{0}

\begin{figure*}[h]
\begin{center}
\includegraphics[width=120 mm,bb= 0 0 394 293]{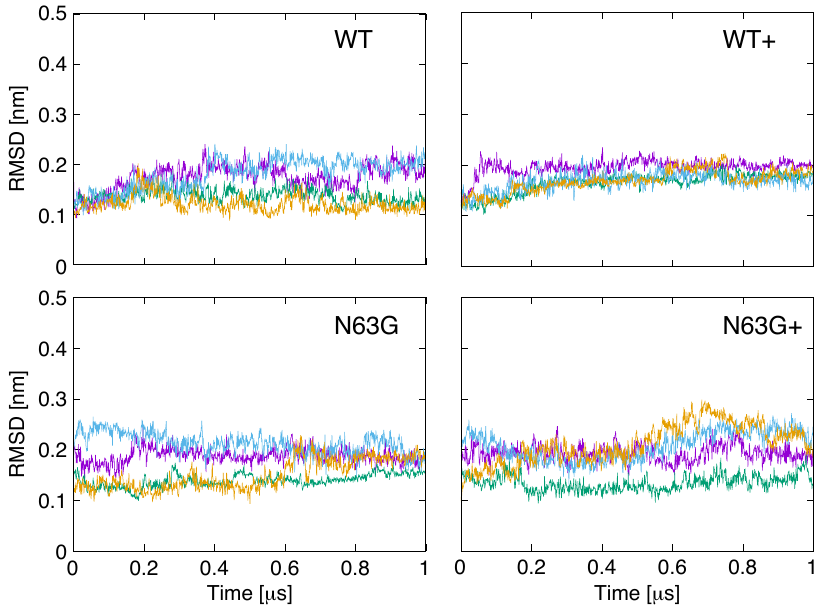}
\caption{Root mean square deviation (RMSD) of AQP6 embedded in a POPC lipid bilayer.
Four different single membrane systems of WT (neutral), WT+ (low pH), N63G (neutral), and N63G+ (low pH) were simulated for $ 1 \, \mathrm{\mu s}$.
RMSD values for each monomer are shown with different colored lines.}
\label{fig_S1}
\end{center}
\end{figure*}

\begin{figure*}[h]
\begin{center}
\includegraphics[width=120 mm,bb= 0 0 272 151]{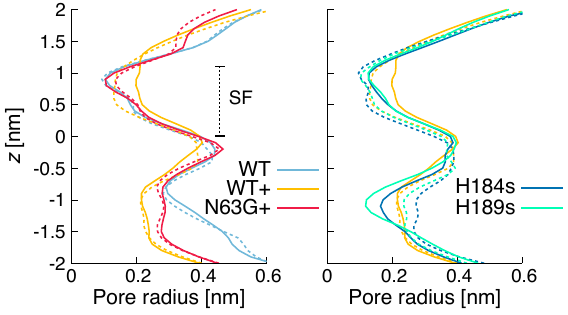}
\caption{Pore radius profile along the central pore in the AQP6 tetramer.
The dashed and solid lines represent the pore radius in the upper and lower membranes, respectively.
$z=0 \, \mathrm{nm}$ corresponds to the position of N63.
The data are averaged over the last $100 \, \mathrm{ns}$ from two separate runs.}
\label{fig_S2}
\end{center}
\end{figure*}

\begin{figure*}[h]
\begin{center}
\includegraphics[width=160 mm,bb= 0 0 866 941]{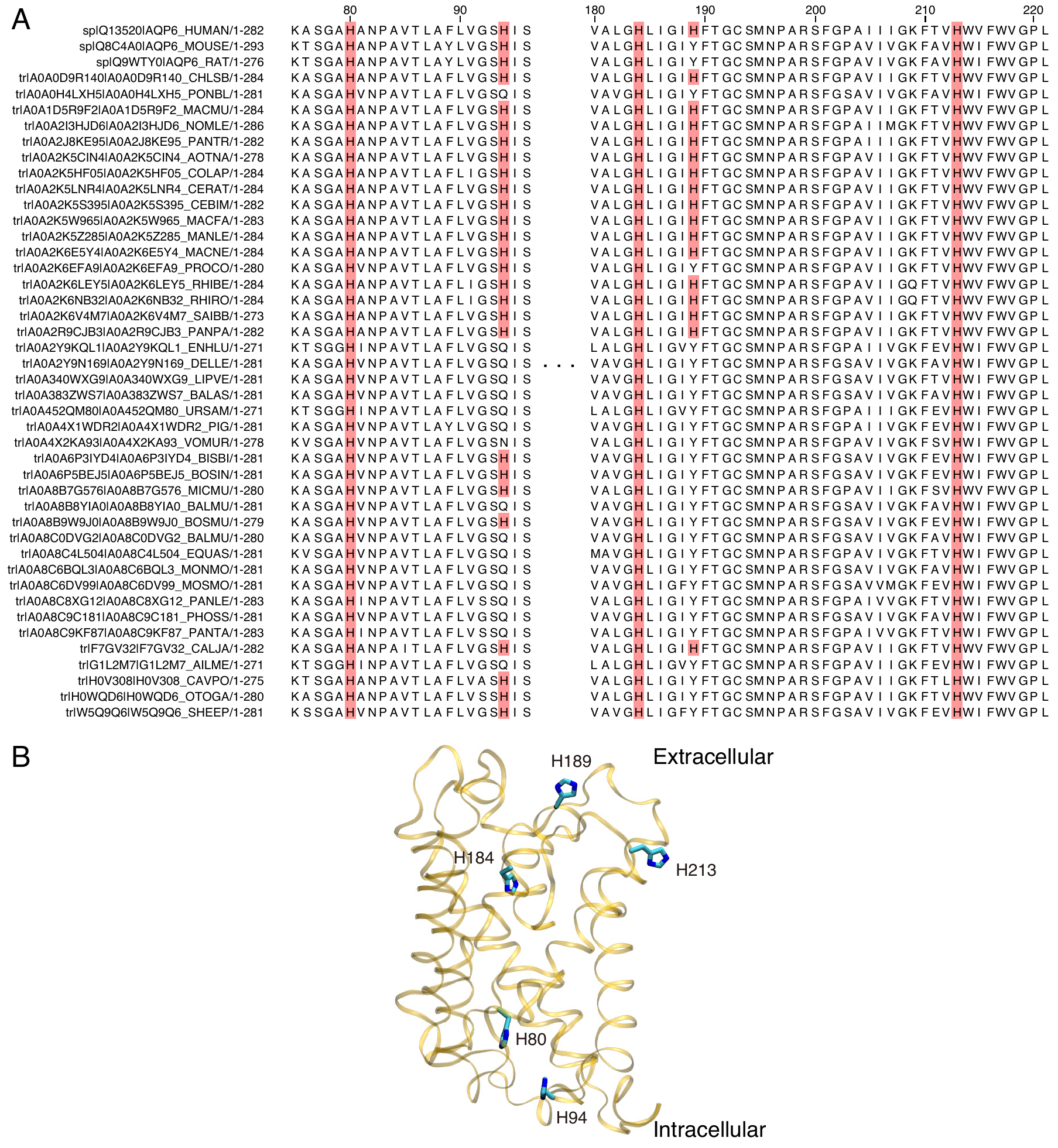}
\caption{Histidine residues in AQP6.
(A)~Sequence comparison of AQP6. Histidine residues are highlighted in red.
(B)~Representative simulation frames of the AQP6 monomer conformation.
Histidine residues are shown in licorice representation.}
\label{fig_S3}
\end{center}
\end{figure*}

\end{document}